# Prediction of nonlinear interface dynamics in the unidirectional freezing of particle suspensions with a rigid compacted layer

Tongxin Zhang, Zhijun Wang∗, Lilin Wang∗,

Junjie Li, Jincheng Wang

*State Key Laboratory of Solidification Processing, Northwestern Polytechnical University, Xi'an 710072, China*

**Abstract:** Ice growth in aqueous suspensions widely exists in natural and industrial settings where freezing of water occurs with a porous medium and results in the spatiotemporal evolution of a solid/liquid interface. Physical models have been proposed in previous efforts to describe the dynamic behaviors of interfaces in the unidirectional freezing of particle suspensions. Most previous models dealt with this process at the single-particle level, and the growth of a particle compacted layer coupled with ice growth was hardly addressed. Here, based on the fundamental momentum theorem, we propose a consistent theoretical framework to address the unidirectional freezing process in particle suspensions coupled with the effect of water permeation. We first propose a constant compacted layer model, in which an interface undercooling-dependent pushing force exerted on the compacted layer with a specific formula is derived based on surface tension. Subsequently, a dynamically growing compacted layer is considered and analysed by a dynamic compacted layer model. Numerical solutions of the nonlinear models reveal the dependence of the system dynamics on some typical physical parameters, such as particle radius, initial particle concentration in the suspensions, and freezing velocity. The system dynamics are characterized by interface velocity, interface undercooling, and interface recoil as functions of time. The models allow us to reconsider the dynamic ice growth during the freezing of particle suspensions in a simple but novel way, with potential implications for both understanding and controlling not only ice formation in porous media but also crystallization processes in other complex systems.

---

∗ Corresponding author. zhjwang@nwpu.edu.cn
∗ Corresponding author. wlilin@nwpu.edu.cn

# I. INTRODUCTION

As one of typical nonlinear and nonequilibrium phenomena in nature, water freezing in particle suspensions is commonly seen in many settings, including cryobiology [1-7], materials science [8-12], cold region science [13, 14] and cryosurgery [15]. The formation pattern of ice morphologies in the freezing of particle suspensions is complex and involves a combination of phase transition, fluid transport, and the thermodynamics of binary solutions. Previous investigations of fluid flow in porous media have described water flow in cases where ice growth occurs simultaneously. Numerous investigations have been based on static analyses of the force balance of a single particle in contact with a growing interface, in which a variety of interactions with specific physical forces occur, ranging from relatively simple criteria to often very sophisticated formulations [16-19]. Although some studies have provided kinetic models [20-24], the role of the compacted layer on water permeation is hardly addressed. In addition, in some recent experimental observations in both particle suspensions [25] and polymer solutions [26], water permeation through a porous compacted layer is suggested to play a predominant role in the growth kinetics of a planar freezing front. In fact, analysis on the single particle level is not directly comparable to the physical process of unidirectional freezing of particle suspensions, where a growing compacted layer coupled with the growing ice in water is present. The permeation relying on the growing compacted layer over time is expected to strongly influence the dynamics of the unidirectional freezing of particle suspensions, which remains further investigation. In a nutshell, different moving interfaces such as ice/water interfaces or particle/water interfaces, are still only partially understood, whether theoretically or experimentally, in the freezing of particle suspensions. Recent analyses of the unidirectional freezing of particle suspensions do not alter skepticism regarding its physical basis.

The aim of this paper is straightforward. Rather than seeking a solution to a single particle-level problem, the physical process of the unidirectional freezing of particle suspensions is addressed by considering the motion of the whole compacted layer based on a simple but widely accepted formulation. The basic formulation of particle/interface interactions was first proposed by Uhlmann et al. [27], which considered the role of a pushing force by surface tension in combination with a viscous drag force. This basic formulation has been successful in predicting many of the experimental results [22, 28-33], which can therefore serve as a basic but reliable formulation in dealing with the physics of particle-interface interactions in our paper. It should be noted that this repulsive force is also equivalently interpreted by van der Waals forces, as suggested by many authors [34-38], by properly choosing a Hamaker constant and a minimum film thickness between particle and solid phase. In this paper, we study the freezing of particle suspensions by focusing on the motion of a compacted layer based on simple but valid assumptions of particle-interface interactions. Here, a pushing force by surface tension together with a viscous drag by water permeation is considered one of the basic components of the interacting force in our model. Based on the basic formulation by Uhlmann et al. [27], we can treat the entire compacted layer of rigid particles as an object in a momentum theorem and propose a simple dynamic model. Our model focuses on the compacted layer of particles directly driven to move in terms of two distinct forces (one being the pushing force $F_R$ and the other being the viscous drag $F_{f_D}$), and no other interactions between the compacted layer and ice are involved. The pushing force $F_R$ is assumed to stem from the solid/liquid surface tension, and the viscous drag $F_{f_D}$ is assumed to be exerted by water inflow governed by Darcy's law. In this basic physical setting, no fitting parameters are involved in our nonlinear models, which nevertheless enable us to capture the transient dynamic behavior of the system and the effects of different variables on it.

## II. REPRESENTATION OF THE MODEL

During unidirectional freezing of water in particle suspensions, particles can be rejected by the ice and accumulate outside of it to form a compacted layer, as shown in **Fig. 1**. Rejection of particles by growing ice forms a porous matrix of compacted layer ahead of ice, which consists of closely packed particles. The compacted layer is formed irreversibly when the intermolecular distance among colloidal particles is reduced to a specific extent, which guarantees a fixed packing density close to random packing [25, 39]. For simplicity, the packing density in the compacted layer is assumed to be constant, and the diffusion of particles is also neglected in our model. Here, we represent the derivation of our model that addresses the transient movement of the solid/particle ($\Gamma_{S/P}$) and the particle/liquid ($\Gamma_{P/L}$) interfaces along with the build-up of the compacted layer with length $L$ during unidirectional freezing of particle suspensions in an elongated squared glass capillary. Our starting point is based on the physical setting shown in **Fig. 1** in the one-dimensional reference frame $Z$, fixed to the ground to present the distance coordinate in the direction of the pulling velocity $V_{pulling}$, whose original point $O$ is chosen as the position of $\Gamma_{S/L}$ for pure water. The unidirectional freezing of the particle suspensions is achieved by the set of a thermal gradient of fixed magnitude $G$ moving horizontally from left to right side at velocity $V_{pulling}$ relative to the frame $Z$. Here, if we choose an isotherm of the freezing point of pure water $T_m$ as an indicator for the movement of $G$, the moving velocity of isotherm $T_m$ is then physically equivalent to the freezing/pulling velocity $V_{pulling}$ of the system. **Fig. 1 (a)** exhibits a constant compacted layer that does not change over time, while **Fig. 1 (b)** is a more realistic case with a dynamically growing compacted layer as a function of time. It should be noted here that the primitive case first discussed in **Fig. 1 (a)** helps us to capture the main features of the modelling before consideration of the dynamics of a more general case with a growing compacted layer in **Fig. 1 (b)**.

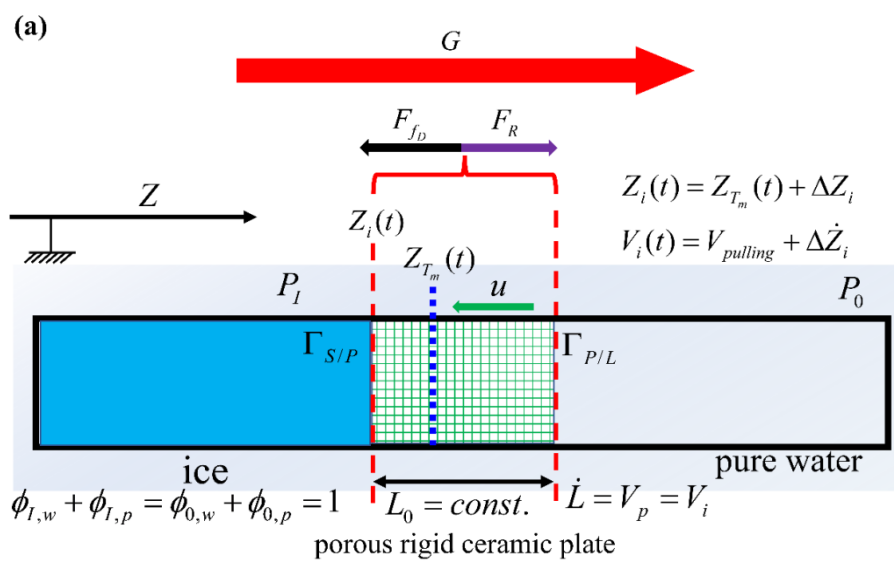

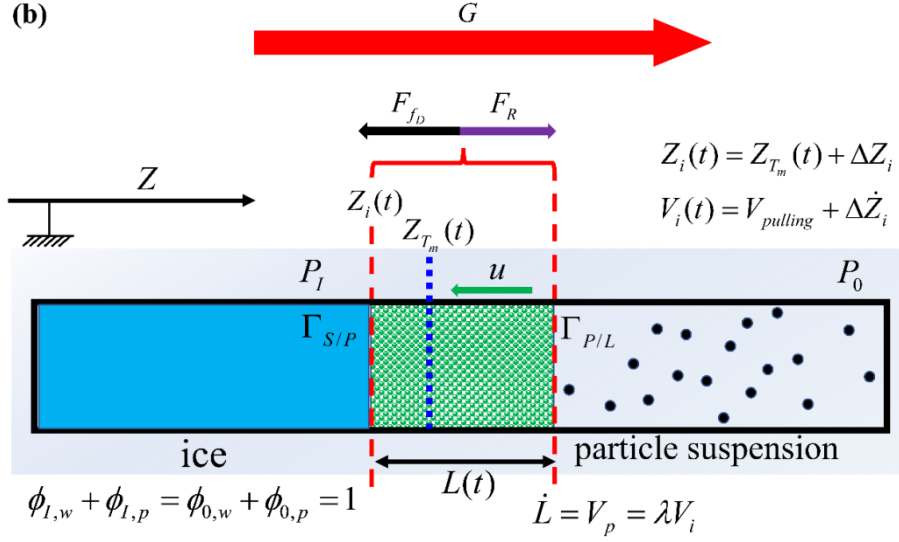

**FIG. 1** Schematic of two cases of unidirectional freezing. The freezing sample is assumed to be static in the frame of reference $Z$, which is fixed to the ground. The isotherm of $T_m$ with its position $Z_{T_m}(t)$ in a thermal gradient of fixed magnitude $G$ moves horizontally from left to right at speed $V_{pulling}$, leading to unidirectional freezing of water in the sample. As the ice grows forwards, the constant compacted layer with a constant packing density $\phi_{I,p}$ of particles is pushed forwards ahead of the ice due to force differences ($F_R$ and $F_{f_D}$) exerted on it to result in a corresponding interface movement $Z_i(t)$. The ice growth is supported by water inflow at a velocity $u$ from right to left through the compacted layer. **(a)** A compacted layer of constant length $L_0$ is placed above an ice/water interface in a glass capillary that is glued onto a large glass sheet, forming two interfaces (the solid/particle interface $\Gamma_{S/P}$ and the particle/liquid interface $\Gamma_{P/L}$) with no particle far from the compacted layer. **(b)** A particle suspension is placed above an ice/water interface in a glass capillary glued onto a large glass sheet. As the ice grows forwards, the particles accumulate ahead of the ice, which results in a dynamically growing compacted layer of particles with a time-dependent length $L(t)$ and a constant packing density of particles $\phi_{I,p}$, forming two interfaces ($\Gamma_{S/P}$ and $\Gamma_{P/L}$).

We consider the conservation of water during ice growth in both cases in **Fig. 1**, as follows. Two distinct phases (water and rigid particles) are related by their volume fractions as

$$\phi_{0,w} + \phi_{0,p} = \phi_{I,w} + \phi_{I,p} = 1 \qquad \text{(Eq. 1)}$$

where $\phi_{0,w}$ is the volume fraction of water far from $\Gamma_{P/L}$, $\phi_{I,w}$ is the volume fraction of water at $\Gamma_{S/P}$, $\phi_{0,p}$ is the volume fraction of particles far from $\Gamma_{P/L}$, $\phi_{I,p}$ is the volume fraction of particles at $\Gamma_{S/P}$, $\rho_{0,w}$ is the volume-averaged density (water mass per unit volume) of water far from $\Gamma_{P/L}$, and $\rho_{I,w}$ is the volume-averaged density of water at $\Gamma_{S/P}$. The growth of ice at velocity $V_i$ is supported by the inflow of water at velocity $u$ from the right to the left of the compacted layer. Following the mass conservation of water over the whole region on the left of $\Gamma_{S/P}$ in the frame $Z$ in an arbitrary time interval $(t_1, t_2)$, we have

$$\rho_{ice} A \int_{t_1}^{t_2} V_i dt = \int_{t_1}^{t_2} dt \iint_{\Gamma_{S/P}} \rho_w u dS \tag{Eq. 2}$$

where $\rho_{ice}$ ($\rho_w$) is the density of pure ice (water). The left-hand side of Eq. 2 is the water mass depleted for ice growth that is characterized by the observed movement of $\Gamma_{S/P}$ in the time interval $(t_1, t_2)$. Because the inner sectional area of the capillary $A$ is known and considering the incompressibility of water ($\rho_w$ = constant), the integral on the right-hand side of Eq. 2 can be easily determined as

$$\int_{t_1}^{t_2} dt \iint_{\Gamma_{S/P}} \rho_w u dS = A \rho_w \int_{t_1}^{t_2} u dt \tag{Eq. 3}$$

Rearranging Eq. 2 by combining it with Eq. 3, we have

$$\int_{t_1}^{t_2} (\rho_{ice} \cdot V_i - \rho_w \cdot u) dt = 0 \tag{Eq. 4}$$

Because the time interval $(t_1, t_2)$ is arbitrary, Eq. 4 yields

$$u = \frac{\rho_{ice}}{\rho_w} V_i \tag{Eq. 5}$$

Based on the mass conservation of water during freezing, Eq. 5 theoretically shows the relation between the water inflow velocity $u$ through the compacted layer needed for ice growth and the ice growth velocity $V_i$ during unidirectional freezing.

In the following sections, we propose physical models to describe the dynamic behavior of the system. To obtain an intuitive physical picture of unidirectional freezing coupled with water permeation through the compacted layer, we first conceive a porous rigid ceramic plate for preliminary analyses, as shown in **section 2.1**. Based on the tactics presented in **section 2.1**, we establish another model for a system with a dynamic compacted layer in **section 2.2**. The numerical solutions of our models are then analysed over a wide range of different parameter values to reveal the underlying dynamics of the system.

**2.1 Model with a constant compacted layer $L_0$**

A porous rigid ceramic plate with a permeability $k$ and a permeation thickness $L_0$ is placed in close contact with an initially planar ice/water interface, and the ceramic plate is assumed to cover the entire surface of the ice, as shown in **Fig. 1 (a)**. This constant compacted layer is assumed to consist of many particles closely glued to each other. The dynamics of the unidirectional freezing system can then be reflected by the dynamics of the ceramic plate incorporated into the system. This case is equivalent to the unidirectional freezing process with a constant compacted layer $L_0$, and no excessive particles are present in the liquid phase ahead. Such treatment is helpful because many physical parameters can be elucidated or proposed in this preliminary case.

In this case, a constant compacted layer $L_0$ is initially placed in close contact with the ice surface and forms two interfaces $\Gamma_{S/P}$ and $\Gamma_{P/L}$, as shown in **Fig. 1 (a)**. Unidirectional freezing starts with this physical setting. Instead of applying a treatment in single particle models, we choose the entire compacted layer in the following analyses. The forces exerted on the compacted layer in our model consist of two parts, namely, the pushing force $F_R$ and the viscous drag $F_{f_D}$ exerted by water flow through the compacted layer, which can be derived from Darcy's law. The dynamics of the compacted layer can be derived based on the momentum theorem if the two forces $F_R$ and $F_{f_D}$ are determined.

On the one hand, the interaction between a solid/liquid interface and insoluble particles mainly arises from surface tension, which results in the requirement of an excessive resistant force to bend or disturb an initially flat interface with solid/liquid surface tension $\gamma_{S/L}$. Correspondingly, another force equal in magnitude but opposite in direction would act on particles in contact with the interface via Newton's Third Law. Accordingly, **Fig. 2 (a)** shows that a compacted layer composed of rigid particles of radius $R_1$ in contact with an initially planar ice/water interface advancing forwards can produce concave indentation and convex pore ice protruding into the porous interstices of neighboring particles. The details of the state of contact at the intersection point of the particle, the ice, and water phases depend on wettability and the thermal conductivities among the three phases and can be expected to vary for different combinations of surface tension and thermal conductivities of the three phases. Here, we consider only the simple case in which the particle is assumed to be nonwettable in the ice phase and the thermal conductivities are identical for the three phases.

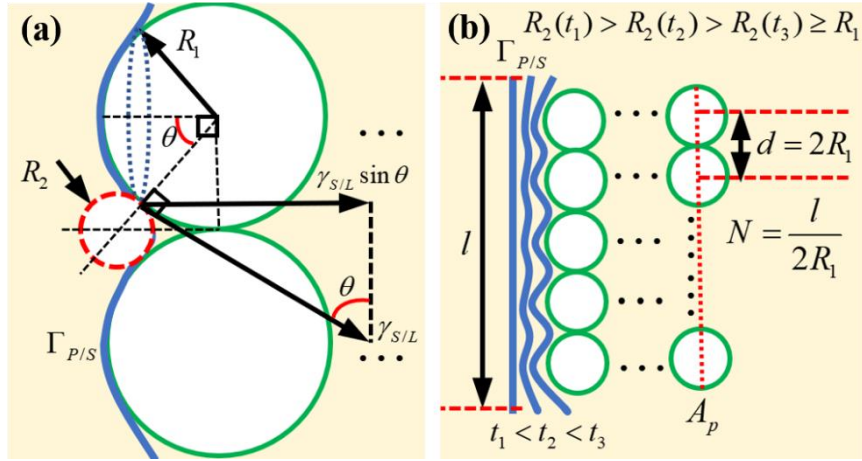

**FIG. 2 (a)** Schematic for the formation of curvature radii for indentation ( $R_1$ ) and pore ice ( $R_2$ ) protruding into the porous interstices of the compacted particles on $\Gamma_{S/P}$ due to its contact with particles with a time-dependent contact angle of $\theta$, which is assumed to satisfy Eq. 6. **(b)** Schematic for the formation of a pushing force exerted on the compacted layer due to surface tension at the ice/water/particle interface with a time-dependent tip radius $R_2(t)$ of pore ice between porous interstices near $\Gamma_{S/P}$, in which $t_1 < t_2 < t_3$, and $R_2(t_1) > R_2(t_2) > R_2(t_3) \geq R_1$.

On the other hand, the curved interface with a curvature radius of $R_2$ at its tip is a free boundary for ice growth without any external force and can be dealt with using a local equilibrium assumption of crystal growth. In thermodynamics, when kinetic undercooling is neglected, the curvature effect of the ice/water interface becomes a major part of its total interface undercooling $\Delta T_i = T_m - T_i$ ($T_i$ is the temperature of the ice/water interface) via the Gibbs-Thompson effect in the form $\Delta T_R = \dfrac{\gamma_{S/L} T_L}{L_m} \cdot \dfrac{1}{R_2}$, and $L_m$ is the released latent heat per unit volume due to phase change from pure water to ice. It is reasonable to assume that the pushing force $F_R$ exerted by the ice/water interface on the compacted layer is dependent on the interface undercooling $F_R = F_R(\Delta T_i)$. We must also find a specific relationship between $\Delta T_i$ and $F_R$.

Here, we can determine the physical form of the pushing force $F_R$ as follows. We first consider the pushing force exerted on a single particle, as indicated in **Fig. 2 (a)**. The exact shape of the convex pore ice growing into the porous interstices of the particle matrix belongs to the results of the dynamic free boundary evolution of the ice/water interface under some physical constraints by neighboring particles that are in contact with it. Rather than resorting to extensive variation principles, we apply a simplified assumption of pore ice with a spherical front geometry. Here, we assume that the front shape of the convex pore ice forms a circle with a radius of $R_2$, as shown in **Fig. 2 (a)**. The circle with radius $R_2$, which overlaps with the pore ice front, is tangent to the two particles with radii $R_1$ on both sides of the particle interstices. The introduction of a contact angle $\theta$ to **Fig. 2 (a)** allows us to easily prove that $R_1$ and $R_2$ are related by

$$(R_2 + R_1)\sin\theta = R_1 \tag{Eq. 6}$$

The contact angle $\theta(t)$ is a time-dependent variable that is zero at the beginning of freezing; it increases over time and affects the magnitude of $R_2$. By taking the possible minimum film thickness $h_0$ between particle and ice, which is comparable to the intermolecular distance $a$ in liquid phase [37, 40, 41], the pushing force of ice exerted on a single contacted particle due to surface tension $F_{Ri}$ is approximated as

$$F_{Ri} = 2\pi R_1 \Delta\gamma \left(\frac{a}{a+h_0}\right)^2 \approx 2\pi R_1 \Delta\gamma \tag{Eq. 7}$$

By further considering the contact area of each particle with ice via a spherical indentation at the intersection of three phases, the pushing force of ice exerted on a single contacted particle $F_{Ri}$ can be linked with the contact angle $\theta$ as

$$F_{Ri} = 2\pi R_1 \sin\theta \cdot [\gamma_{S/L}\sin\theta + (\gamma_{P/L} - \gamma_{P/S})\sin\theta] \approx 2\pi R_1 \sin\theta \cdot \gamma_{S/L}\sin\theta \tag{Eq. 8}$$

where $\gamma_{S/L}$, $\gamma_{P/L}$ and $\gamma_{P/S}$ are the surface tensions of the ice/water interface, the particle/water interface, and the particle/ice interface, respectively. In Eq. 7, it is assumed that the difference in surface tension $\gamma_{P/L} - \gamma_{P/S}$ is negligibly small compared to $\gamma_{S/L}$. Thus, the total pushing force exerted on all particles in contact with the ice surface is

$$F_R = \sum_{i=1}^{N} F_{Ri} = 2N\pi R_1 \gamma_{S/L} \sin^2\theta \leq 2N\pi R_1 \gamma_{S/L} \tag{Eq. 9}$$

where $N$ is the total number of particles in contact with the ice, which simply equals the interface length $l$ divided by the particle diameter $2R_1$, as shown in **Fig. 2 (b)**. It can be seen in Eq. 8 that the maximum value of pushing force $F_R$ is $F_{R\max} = 2N\pi R_1 \gamma_{S/L}$, indicating that the pushing force $F_R(\Delta T_i)$ cannot increase indefinitely with $\Delta T_i$. At this point, $F_R$ is still not related to $\Delta T_i$. Here, by replacing $R_2$ with $\Delta T_R$ and using the approximation of $\Delta T_i \approx \Delta T_R$ in combination with Eq. 6 and Eq. 8, we can obtain a functional form of $F_R(\Delta T_i)$ that appears adequate for our purposes in Eq. 9 as

$$F_R(\Delta T_i) = F_{R\max}(1 - \frac{1}{1 + \frac{R_1 L_m}{\gamma_{S/L} T_m}\Delta T_R})^2 \leq F_{R\max} \tag{Eq. 10}$$

A key property of $F_R(\Delta T_i)$ in Eq. 9 is that it has the theoretically limiting value $F_{R\max}$. However, it can be shown that $\Delta T_R$ cannot increase infinitely, which limits the ability of $F_R(\Delta T_i)$ to increase. In the two-dimensional case examined in this paper, the curvature radius of pore ice $R_2$ can decrease indefinitely, approaching zero with a corresponding infinitely large $\Delta T_R$, but this never occurs because in three-dimensional cases, the minimum curvature radius $R_{2\min}$ is determined by the void space formed by the three particles in close contact with each other. Using a geometrical relationship between a circle at the center of the void that is tangent to the three particles, it is easy to prove that the ratio of $\frac{R_{2\min}}{R_1}$ is $\frac{2-\sqrt{3}}{\sqrt{3}}$. This ratio requires a more realistic maximum curvature undercooling of $\Delta T_{R\max} = \frac{\gamma_{S/L} T_m}{L_m R_{2\min}}$, after which the pore ice/ice spears are expected to develop and protrude deeply into the porous interstices of the compacted layer. Thus, we can define *the critical time $t_c$* as the time $t$ when $\Delta T_R$ reaches *the critical interface undercooling $\Delta T_{R\max}$*.

In addition to pushing force $F_R(\Delta T_i)$, we need to determine the viscous drag $F_{f_D}$ exerted on the compacted layer. It has been shown [42] that the basic Navier-Stokes equation for a steady fluid flow at low Reynolds numbers ($\text{Re} \ll 1$) in a uniform porous medium can be reduced to the well-known Darcy's law by neglecting the inertial and form drag terms. Thus, the main force that the water permeation flow exerts on the compacted layer is viscous drag $F_{f_D}$. The water inflow through a homogeneously porous compacted layer supports ice growth during unidirectional freezing in a form governed by Darcy's law at low Reynolds numbers ($\text{Re} \ll 1$ in this paper). Darcy's law for water permeation (at an area-averaged velocity $u$ with respect to a sectional area of $A$) relative to the center of mass of the moving compacted layer (at velocity $V_i$) in frame $Z$ satisfies

$$u + V_i = \frac{Q}{A} = \frac{k(\phi_p, r)}{\mu} \frac{dP}{dL_0} \approx \frac{k(\phi_p, r)}{\mu} \frac{\Delta P}{L_0} \quad (\text{Eq. 11})$$

where $\Delta P = P_0 - P_I \in (0, P_0)$ is the Darcy pressure difference on both sides of $\Gamma_{S/P}$, which drives water to permeate from far from $\Gamma_{P/L}$ to $\Gamma_{S/P}$, $Q$ is the volume flux of water per unit time through the compact layer, $A$ is the inner sectional area of the capillary, $k(\phi_p, R_1)$ is the permeability of the compact layer that depends on both the volume fraction $\phi_p$ and the particle radius $R_1$ of the particle matrix in the compacted layer, according to the well-known Kozeny–Carman equation [43], as $k(\phi_p, R_1) = \frac{R_1^2 (1-\phi_p)^3}{45 \phi_p^2}$, and $\mu$ is the viscosity of pure water as permeation fluid. By multiplying the pressure difference term $\Delta P$ in Eq. 11 by the total sectional area $A$ of the compacted layer, the viscous drag $F_{f_D}$ exerted by water permeation flow on the compacted layer can be proven to satisfy

$$F_{f_D} = \min\{\frac{A\mu L_0}{k} \cdot (u + V_i), F_{f_D \max}\} \in (0, AP_0) \quad (\text{Eq. 12})$$

where $F_{f_D \max} = AP_0$ is the maximum viscous drag. Eq. 12 indicates that $F_{f_D}$ is physically limited by a maximum value because the Darcy pressure at $\Gamma_{S/P}$ can only be larger than zero, which serves as a boundary condition for Darcy pressure, as well as limited water permeation. The limited water flow determines the growth velocity of the solid/liquid interface, which may be smaller than the pulling velocity and induces a continuous recoil of $\Gamma_{S/P}$ relative to the position of isotherm $T_m$ in the frame $Z$.

From the above analysis, we obtain the force components exerted on the compacted layer. In the following, we derive the dynamics of the system by focusing on the momentum theorem of the compacted layer. By further neglecting the effects of gravity as a force for the compacted layer, the movement of the compacted layer of mass $m_p$ can be simply addressed by the force differences exerted on it. Its momentum theorem in the frame $Z$ gives

$$F_R - F_{f_D} = \frac{dP_p}{dt} = m_p \frac{dV_{com}}{dt} \qquad \text{(Eq. 13)}$$

where $m_p = \rho_{I,p} A L_0$ is the mass of all particles in the constant compacted layer of length $L_0$, $P_p = m_p V_{com}$ is the momentum of all particles in the compacted layer, and $V_{com}$ is the velocity of the center of mass of all particles in the compacted layer. Here, $V_{com}$ simply equals $V_i$ for a constant compacted layer of rigid particles whose length does not grow with $t$. The position of $\Gamma_{S/P}$ is given as $Z_i(t)$, which is related to the position of the isotherm line $Z_{T_m}(t)$ as

$$Z_i(t) = Z_{T_m}(t) + \Delta Z_i(t) \qquad \text{(Eq. 14)}$$

where $\Delta Z_i(t)$ is the recoil of $\Gamma_{S/P}$ with respect to $Z_{T_m}(t)$. Differentiating Eq. 14 with respect to time $t$ yields

$$V_i(t) = V_{pulling} + \Delta \dot{Z}_i(t) = V_{pulling} - \frac{\Delta \dot{T}_i}{G} \qquad \text{(Eq. 15)}$$

Combining Eqs. 10–12, 13, and 15, we find that the governing nonlinear ordinary differential equations (ODEs) describing the dynamic evolution of the unidirectional freezing system with a constant compacted layer can be derived as follows:

$$\frac{d}{dt}\begin{bmatrix} V_i \\ \Delta T_i \end{bmatrix} = \begin{bmatrix} [F_R(\Delta T_i) - F_{f_D}(V_i)]/(\rho_{I,p} A L_0) \\ G \cdot (V_{pulling} - V_i) \end{bmatrix} \qquad \text{(Eq. 16)}$$

with the initial conditions

$$\begin{bmatrix} V_i \\ \Delta T_i \end{bmatrix}\bigg|_{t=0} = \begin{bmatrix} 0 \\ 0 \end{bmatrix} \qquad \text{(Eq. 17)}$$

The ODEs in Eq. 16 can be applied to reveal the dynamic information of the system with a constant compacted layer.

**2.2 Model with a dynamic compacted layer $L(t)$**

By analogy to the analysis given in section **2.1**, this section addresses a model with a dynamic compacted layer whose length $L(t)$ grows over time $t$. In this case, the dynamic compacted layer $L(t)$ is considered, which also forms two interfaces $\Gamma_{S/P}$ and $\Gamma_{P/L}$, as shown in **Fig. 1 (b)**. A unidirectional freezing process begins with this physical setting. The mass conservation of particles around the compacted layer in a fixed region includes the compacted layer in the frame $Z$, which proves that the growth velocities of the compacted layer $\dot{L}$ (equals the velocity of $\Gamma_{P/L}$ in the frame $Z$) and ice $V_i$ (i.e., the velocity of $\Gamma_{S/P}$ in the frame $Z$) are related by

$$\dot{L} = V_p = \lambda \cdot V_i \tag{Eq. 18}$$

where $\lambda = \dfrac{\rho_{0,p}}{\rho_{I,p} - \rho_{0,p}}$ depends on the volume-averaged density of particles in the compact layer region $\rho_{I,p}$ and in the liquid phase $\rho_{0,p}$ far from $\Gamma_{P/L}$. Eq. 18 shows that the movement of $\Gamma_{P/L}$ and $\Gamma_{S/P}$ can be obtained if one of them is determined. Similar to the tactics in **section 2.1**, the dynamics of $\Gamma_{P/L}$ can be reflected by choosing the whole dynamic compacted layer as an object for analysis. The momentum theorem of a dynamic compacted layer employs a more general form:

$$F_R - F_{f_D} = \frac{dP_p}{dt} = m_p \frac{dV_{com}}{dt} + V_{com} \frac{dm_p}{dt} \tag{Eq. 19}$$

where $V_{com}$ is the velocity of the center of mass of the dynamic compacted layer in frame $Z$ and $\dfrac{dm_p}{dt}$ is the variation rate of the total particle mass in the dynamic compacted layer. In this case, $L(t)$ increases with $t$ due to the condensation of particles from dilute particle suspensions far from $\Gamma_{P/L}$. Due to the increase in $L(t)$ as a function of time $t$, the variation rate of the total particle mass in compact layer $\frac{dm_p}{dt}$ can be determined as follows. The mass variation $dm_p$ in time interval $dt$ is

$$dm_p = \rho_{I,p} A dL \tag{Eq. 20}$$

By differentiating Eq. 20 with respect to time $t$, we obtain

$$\frac{dm_p}{dt} = \rho_{I,p} A \cdot \dot{L} \tag{Eq. 21}$$

Because $L(t)$ increases over time $t$, the center of mass of the dynamic compacted layer moves with a velocity $V_{com}$ larger than $V_i$ in frame $Z$, which can be easily proven to be

$$V_{com} = V_i + \frac{1}{2}\dot{L} = (\frac{1}{\lambda} + \frac{1}{2})\dot{L} \tag{Eq. 22}$$

Differentiating Eq. 22 with respect to time $t$, we obtain

$$\frac{dV_{com}}{dt} = (\frac{1}{\lambda} + \frac{1}{2})\ddot{L} \tag{Eq. 23}$$

Note that for the dynamic compacted layer here, its $F_R$ takes the same form as that in Eq. 10, but its $F_{f_D}$ becomes slightly different from that in Eq. 12. Analogous to Eq. 10 and recalling the time-dependent property of $L(t)$, $F_{f_D}$ is now a function of both $L$ and $\dot{L}$, which adopts a similar form in Eq. 10 combined with Eq. 18, as

$$\begin{aligned} F_{f_D}(L,\dot{L}) &= \min\{\frac{A\mu L}{k} \cdot (u + V_{com}), F_{f_{D\max}}\} \\ &= \min\{\frac{A\mu(\rho_{ice}/\rho_w + 1 + \lambda/2)}{k\lambda} \cdot \dot{L} \cdot L, F_{f_{D\max}}\} \in (0, A_w P_0) \end{aligned} \tag{Eq. 24}$$

When Eqs. 18, 19, and 21–24 are combined, the governing nonlinear ODEs describing the dynamic evolution for the unidirectional freezing system with a dynamic compacted layer can be derived as

$$\frac{d}{dt}\begin{bmatrix} L \\ \dot{L} \\ \Delta T_i \end{bmatrix} = \begin{bmatrix} \dot{L} \\ \dfrac{\lambda}{\rho_{I,p} A} \cdot \dfrac{F_R(\Delta T_i)}{L} - \dfrac{\lambda}{\rho_{I,p} A} \cdot \dfrac{F_{f_D}(L,\dot{L})}{L} - (1+\dfrac{\lambda}{2}) \cdot \dfrac{\dot{L}^2}{L} \\ G \cdot (V_{pulling} - \dfrac{\dot{L}}{\lambda}) \end{bmatrix} \qquad \text{(Eq. 25)}$$

with the initial conditions

$$\begin{bmatrix} L \\ \dot{L} \\ \Delta T_i \end{bmatrix}\Bigg|_{t=0} = \begin{bmatrix} \tilde{R}_1 \\ 0 \\ 0 \end{bmatrix} \qquad \text{(Eq. 26)}$$

Here, the symbol $\tilde{R}_1$ in Eq. 26 signifies that the initial value of $L$ for the numerical solution should be given in the same order as that of particle radius $R_1$ because the compacted layer grows layer by layer and much larger or smaller initial values for $L$ would be physically unrealistic.

### III. RESULTS AND DISCUSSION

This section presents a series of numerical solutions and relevant discussions of the two derived models. We investigated several combinations of typical parameters that are frequently chosen as key parameters in relevant investigations to explore their possible effects on the nonlinear dynamic behavior of the unidirectional freezing system. In addition, the total interface recoil $\Delta Z_i(t)$ due to limited water permeation through the compacted layer is also characterized as the recoil of $\Gamma_{S/P}$ with respect to isotherm $T_m$ in frame $Z$ as

$$\Delta Z_i(t) = \int_0^t (V_I - V_{pulling}) dt' \leq 0 \qquad \text{(Eq. 27)}$$

Here, the $\Delta Z_i(t)$ as a function of time $t$ is calculated based on the numerical results of $V_I$ in our models. In addition, the critical time $t_c(\Delta T_{R\max})$ is determined for the numerical results for all combinations of physical parameters.

The systems of the two models above are made nondimensional with the following scales. For the model with a constant compacted layer: time $[t] = L_0 / V_{pulling}$ ; temperature $[T] = \gamma_{S/L} T_L / L_m R_1$ ; velocity $[V] = k F_{R\max} / A\mu L_0 (\rho_{ice}/\rho_w + 1)$. For the model with a dynamically growing compacted layer: time $[t] = R_1 / V_{pulling}$ ; temperature $[T] = \gamma_{S/L} T_L / L_m R_1$ ; velocity $[\dot{L}] = k\lambda F_{R\max} / A\mu R_{\text{l}}(\rho_{ice} / \rho_w + \lambda / 2 + 1)$. Indicating rescaled quantities with bars, we first define nondimensional variables for the model with a constant compacted layer

$$\bar{t} = \frac{t}{[t]}, \quad \Delta\bar{T}_i = \frac{\Delta T_i}{[T]}, \quad \bar{V}_i = \frac{V_i}{[V]} \tag{Eq. 28}$$

Applying this scaling to Eq. 16-17 gives the nondimensional ODEs below for the model with a constant compacted layer

$$\frac{d}{d\bar{t}}\begin{bmatrix} \bar{V}_i \\ \Delta\bar{T}_i \end{bmatrix} = \begin{bmatrix} [(1 - \frac{1}{1+\Delta\bar{T}_i})^2 - \min\{\bar{V}_i, \frac{F_{f_D\max}}{F_{R\max}}\}]/C_1 \\ C_2 \cdot (\frac{V_{pulling}}{[V]} - \bar{V}_i) \end{bmatrix} \tag{Eq. 29}$$

where the constants $C_1$ and $C_2$ satisfy

$$C_1 = \frac{\rho_{I,p} A L_0 [V]}{F_{R\max}[t]}, \quad C_2 = \frac{G[V][t]}{[T]} \tag{Eq. 30}$$

Similarly, we then define nondimensional variables for the model with a dynamically growing compacted layer

$$\bar{t} = \frac{t}{[t]}, \quad \Delta\bar{T}_i = \frac{\Delta T_i}{[T]}, \quad \bar{\dot{L}} = \frac{\dot{L}}{[\dot{L}]} \tag{Eq. 31}$$

Applying this scaling to Eq. 25-26 gives the nondimensional ODEs below for the model with a dynamically growing compacted layer.

$$\frac{d}{d\bar{t}}\begin{bmatrix} \bar{L} \\ \bar{\dot{L}} \\ \Delta\bar{T}_i \end{bmatrix} = \begin{bmatrix} \bar{\dot{L}} \\ [(1 - \frac{1}{1+\Delta\bar{T}_i})^2 - \min\{\frac{[L]}{R_1}\bar{L}\bar{\dot{L}}, \frac{F_{f_D\max}}{F_{R\max}}\} - C_4 \cdot \bar{\dot{L}}^2]/C_3\bar{L} \\ C_5 \cdot (\frac{V_{pulling}}{[\dot{L}]} - \frac{\bar{\dot{L}}}{\lambda}) \end{bmatrix} \tag{Eq. 32}$$

where the constants $C_3$, $C_4$, and $C_5$ satisfy

$$C_3 = \frac{\rho_{I,p} A(1/\lambda + 1/2)[L][\ddot{L}]}{F_{R\max}}, \quad C_4 = \frac{\rho_{I,p} A(1/\lambda + 1/2)[\dot{L}]^2}{F_{R\max}}, \quad C_5 = \frac{G[\dot{L}][t]}{[T]} \tag{Eq. 33}$$

### 3.1 Results for a model with a constant compacted layer $L_0$

We know of no exact analytic solution of the set of nonlinear ODEs in Eq. 29, and the numerical solutions to Eq. 29 are considered in this paper. Nevertheless, it is shown in detail in **Appendix A** that by neglecting the term $d\bar{V}_i/d\bar{t}$ (the inertia of the compacted layer), an exact solution can be found, which reproduces very close results to the numerical solution of Eq. 29. Here, a standard setting of physical parameters includes $V_{pulling}$ = 2 μm/s, $R_1$ =1000 nm, $L_0$ = 10 μm, and $G$ = 3 K/mm, with other parameters fixed as provided in **Table I**.

It is guaranteed that when we vary any one of $V_{pulling}$, $R_1$, $L_0$, and $G$, other parameters are made to equal those in the standard setting. The choice of parameters in **Table I** leads to the numerical solution of Eq. 29. Here, we tested four typical physical parameters ($V_{pulling}$, $R_1$, $L_0$, and $G$) in our model with a constant compacted layer to explore their possible effects on the system dynamics. From the numerical results, the system dynamics are characterized by three variables $\bar{V}_i$, $\Delta \bar{T}_i$, and $\Delta \bar{Z}_i$.

**TABLE I** Physical parameters utilized for numerical solutions in the model with a constant compacted layer. The parameters in a standard setting are labelled in red. The time step is taken to be 0.1 s.

| Parameters | Value | Units *(SI)* |
|---|---|---|
| $V_{pulling}$ | 1, **2**, 6, 8 | $10^{-6}$ m/s |
| $R_1$ | 300, 500, **1000**, 3000 | $10^{-9}$ m |
| $\phi_{I,p}$ | 0.7 | 1 |
| $\rho_1$ | 1.0 | $10^3$ kg/m$^3$ |
| $\rho_p$ | 1.1 | $10^3$ kg/m$^3$ |
| $\mu$ | 1.7921 | $10^{-3}$ Pa·s |
| $p_0$ | 1.01 | $10^5$ Pa |
| $\gamma_{S/L}$ | 35 | $10^{-3}$ J/m$^2$ |
| $L_m$ | 3.06 | $10^8$ J/m$^3$ |
| $G$ | 1, **3**, 5, 7 | $10^3$ K/m |
| $L_0$ | **10**, 100, 300, 400 | $10^{-6}$ m |
| $l$ | 1 | $10^{-3}$ m |

**Figure 3** shows the effects of different $V_{pulling}$ on system dynamics with other fixed parameters in **Table I**. In the case of **Fig. 3**, $V_{pulling}$ controls the system dynamics via its relation with $\Delta \bar{T}_i$, which affects $F_R$. In **Fig. 3 (a)**, two of the interface velocities $\bar{V}_i$ first undergo a transient acceleration stage and asymptotically approach a steady state, with the other two reaching their $t_c$ earlier than a steady state. Under this steady state, $\Delta \bar{T}_i$ and $\Delta \bar{Z}_i$ also remain unchanged, as shown in **Fig. 3 (b)** and **Fig. 3 (d)**, respectively. Almost no difference is seen among the trajectories of different levels of $V_{pulling}$ in the ($\bar{V}_i, \Delta \bar{T}_i$) plane, as shown in **Fig. 3 (c)**, which indicates that $V_{pulling}$ can only accelerate the system dynamics but will not yield a differentiated dynamic path.

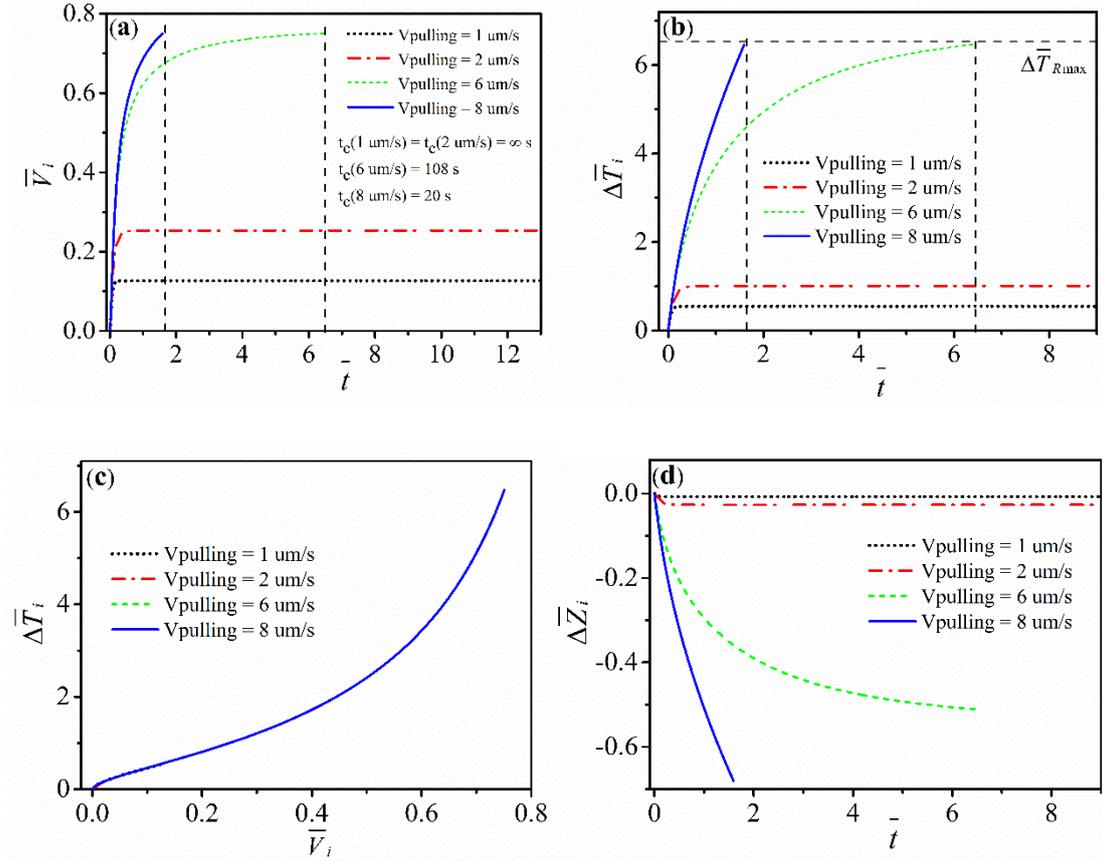

**FIG. 3** Effects of different levels of $V_{pulling}$ on system dynamics with other fixed parameters from **Table I** for our model with a constant compacted layer. **(a)** $\bar{V}_i$-$\bar{t}$ curve with different $t_c$ = 108, and 20 s for $V_{pulling}$ = 6, and 8 μm/s, respectively, the $t_c$ for $V_{pulling}$ = 1, and 2 μm/s being infinite due to the appearance of their steady state before reaching their $\Delta \bar{T}_{R\max}$, the vertical dashed lines indicating the physical bound by the $t_c$; **(b)** $\Delta \bar{T}_i$ - $\bar{t}$ curve with labelled $\Delta \bar{T}_{R\max}$ and $t_c$ in dashed lines; **(c)** Trajectory in the ($\bar{V}_i$, $\Delta \bar{T}_i$) plane; **(d)** $\Delta \bar{Z}_i$ - $\bar{t}$ curve.

The observed steady state in **Fig. 3** corresponds to the physical state when the total pushing force equilibrates the viscous drag force exerted on the compacted layer. In single particle models [23, 24, 44, 45], a similar steady state also exists, in which a particle gradually approaches the growing interface and accelerates, resulting in an increase in both pushing force and viscous drag force. After a certain period of time, the distance between the particle and interface remains constant [23, 44], and the particle is pushed ahead at a constant velocity. Under such a steady state, the pushing force equilibrates the drag force for a particle interacting with the interface. Here, for the results in our constant compacted layer model, it is therefore reasonable to expect a similar steady state since a constant compacted layer can be regarded as a large "particle" interacting with the interface.

However, it is further revealed by our numerical calculation under different sets of parameters that this steady state can only exist under some physical constraints. It should be noted that this steady state is physically possible only before the critical time is reached. As defined earlier, the critical time $t_c$ is the time when the interface undercooling reaches a critical value to allow further development of pore ice into the pore interstices of the compacted layer. If the interface velocity $\bar{V}_i$ can accelerate to keep up with the pulling velocity $V_{pulling}$ earlier than $t_c$, $\Delta \bar{Z}_i$ will reach a steady state, resulting in a steady state of $\Delta \bar{T}_i$ lower than $\Delta \bar{T}_{R\max}$. In this case, critical undercooling will never be reached, and a steady state will rigorously exist over time. This situation usually occurs for compacted layers with higher permeability (i.e., larger $R_1$, as shown in **Fig. 4 (a)**, and smaller $L_0$, as shown in **Fig. 5 (a)**) and under lower $V_{pulling}$, as shown in **Fig. 3 (a)**. On the other hand, if the interface kinetics accelerates to a very limited extent below $V_{pulling}$, the interface recoil will increase to reach the critical interface undercooling, and the critical time will be reached before an observable steady state appears, resulting in no possible steady state over time. This usually occurs for the compacted layers with lower permeability (i.e., smaller $R_1$, as shown in **Fig. 4 (a)**, and larger $L_0$, as shown in **Fig. 5 (a)**) and under higher $V_{pulling}$, as shown in **Fig. 3 (a)**. The freezing process enters the regime of ice spear growth when the critical time is reached, which is beyond the scope of our model and serves as a *physical bound* for $\bar{V}_i$, $\Delta \bar{T}_i$ and $\Delta \bar{Z}_i$. The calculated data points for all quantities beyond the range of critical time are thus *invalid* and not included in our results.

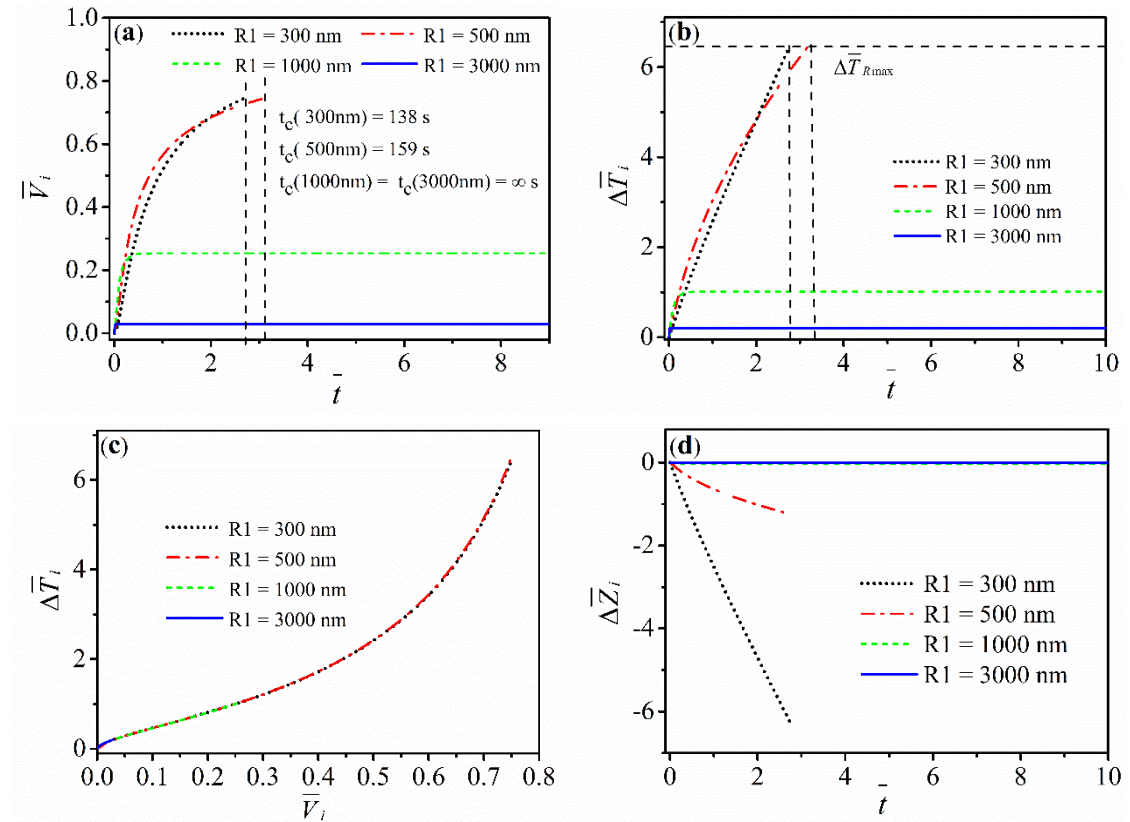

**FIG. 4** Effects of different levels of $R_1$ on system dynamics with other fixed parameters from **Table I** for our model with a constant compacted layer. **(a)** $\bar{V}_i$ - $\bar{t}$ curve with different critical times $t_c$ = 138, and 159 s for $R_1$ = 300, and 500 nm, respectively, the $t_c$ for $R_1$ = 1000,

and 3000 nm being infinite due to the appearance of their steady state before reaching their $\Delta \bar{T}_{R\max}$, the vertical dashed lines indicating the physical bound by the $t_c$; **(b)** $\Delta \bar{T}_i$ - $\bar{t}$ curve with labelled $\Delta \bar{T}_{R\max}$ and $t_c$ in dashed lines; **(c)** Trajectory in the ($\bar{V}_i$, $\Delta \bar{T}_i$) plane; **(d)** $\Delta \bar{Z}_i$ - $\bar{t}$ curve.

**Figure 4** shows the effects of different $R_1$ values of 300, 500, 1000, and 3000 nm on the system dynamics with other fixed parameters in **Table I**. In the case of **Fig. 4**, $R_1$ controls the system dynamics in a more complex manner via its influence on both the $k$ of the compacted layer and the particle-ice interaction, which affects both $F_{f_D}$ and $F_R$. In **Fig. 4 (a)**, two of the interface velocities $\bar{V}_i$ first undergo a transient acceleration stage and asymptotically approach a steady state, with the other two reaching their $t_c$ earlier than a steady state. Under this steady state, $\Delta \bar{T}_i$ and $\Delta \bar{Z}_i$ also remain unchanged, as shown in **Fig. 4 (b)** and **Fig. 4 (d)**, respectively. Different values for $R_1$ produce nearly the same trajectories in the ($\bar{V}_i$, $\Delta \bar{T}_i$) plane, as shown in **Fig. 4 (c)**.

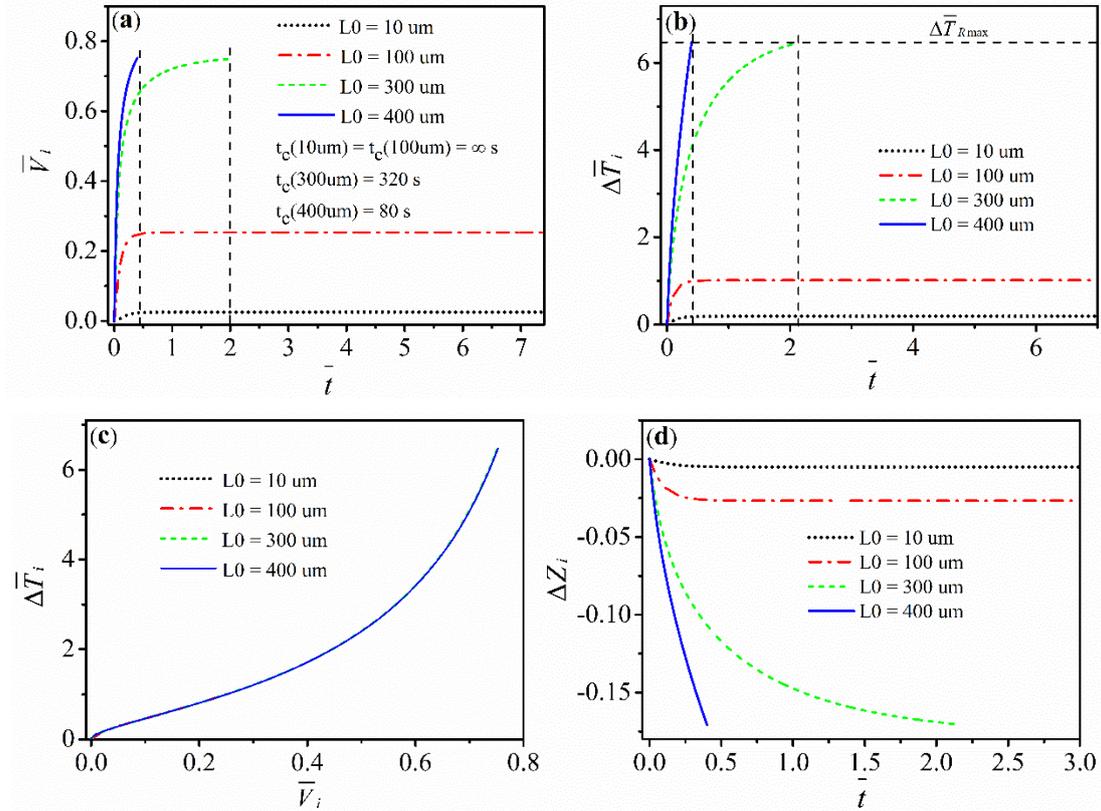

**FIG. 5** Effect of different levels of $L_0$ on the system dynamics with other fixed parameters from **Table I** for our model with a constant compacted layer. **(a)** $\bar{V}_i$ - $\bar{t}$ curve with roughly the same critical time $t_c$ = 80, and 320 s for $L_0$ = 300, and 400 μm, respectively, the $t_c$ for $L_0$ = 10, and 100 μm being infinite due to the appearance of their steady state before reaching their $\Delta \overline{T}_{R\max}$, the vertical dashed lines indicating the physical bound by the $t_c$; **(b)** $\Delta \overline{T}_i$-$\overline{t}$ curve with labelled $\Delta \overline{T}_{R\max}$ and $t_c$ in dashed lines; **(c)** Trajectory in the ($\overline{V}_i$, $\Delta \overline{T}_i$) plane; **(d)** $\Delta \overline{Z}_i$-$\overline{t}$ curve.

**Figure 5** shows the effects of different $L_0$ values of 10, 100, 300, and 400 μm on the system dynamics with other fixed parameters in **Table I**. In the case of **Fig. 5**, $L_0$ controls the system dynamics via its relation with $F_{f_D}$. In **Fig. 5 (a)**, two of the interface velocities $\overline{V}_i$ first undergo a transient acceleration stage and asymptotically approach a steady state, with the other two reaching their $t_c$ earlier than a steady state. Under this steady state, $\Delta \overline{T}_i$ and $\Delta \overline{Z}_i$ also remain unchanged, as shown in **Fig. 5 (b)** and **Fig. 5 (d)**, respectively. Different values of $L_0$ produce almost the same trajectories in the ($\overline{V}_i$, $\Delta \overline{T}_i$) plane, as shown in **Fig. 5 (c)**.

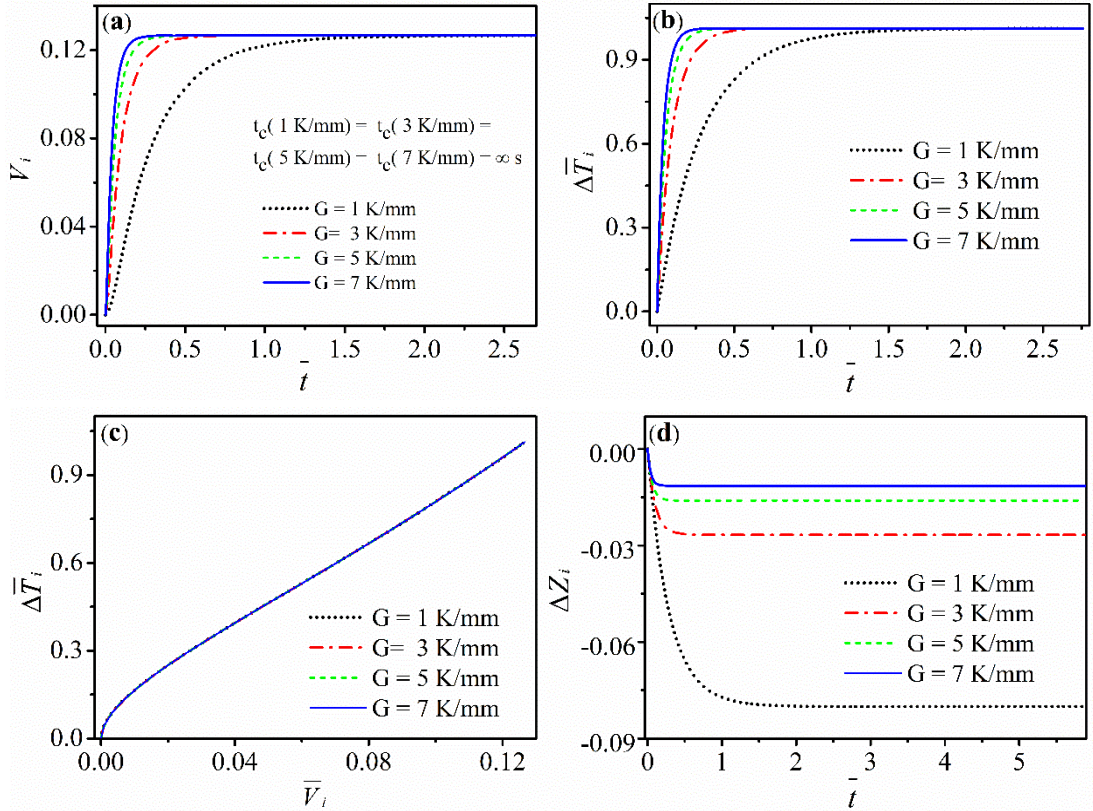

**FIG. 6** Effects of different levels of $G$ on the system dynamics with other fixed parameters from **Table I** for our model with a constant compacted layer. **(a)** $\overline{V}_i$-$\overline{t}$ curve with infinite critical

times $t_c$ for $G$ = 1, 3, 5, and 7 K/mm due to the appearance of their steady state before reaching their $\Delta \overline{T}_{R\max}$; **(b)** $\Delta \overline{T}_i$-$\overline{t}$ curve with labelled $\Delta \overline{T}_{R\max}$ in dashed line; **(c)** Trajectory in the ($\overline{V}_i$, $\Delta \overline{T}_i$) plane; **(d)** $\Delta \overline{Z}_i$-$\overline{t}$ curve.

**Figure 6** shows the effects of different $G$ values of 1, 3, 5, and 7 K/mm on the system dynamics with other fixed parameters in **Table I**. In the cases given in **Fig. 6**, $G$ controls the system dynamics in a simple manner via its relation with $\Delta \overline{T}_i$, which affects $\overline{F}_R$. In **Fig. 6 (a)**, all interface velocities $\overline{V}_i$ first undergo a transient acceleration stage and asymptotically approach a steady state. Under this steady state, $\Delta \overline{T}_i$ and $\Delta \overline{Z}_i$ also remain unchanged, as shown in **Fig. 6 (b)** and **Fig. 6 (d)**, respectively. No distinct difference is seen among the trajectories of different levels of $G$ in the ($\overline{V}_i, \Delta \overline{T}_i$) plane, as shown in **Fig. 6 (c)**, which indicates that $G$ does not yield a differentiated dynamic path.

### 3.2 Results for the model with a dynamic compacted layer $L(t)$

In this section, using proper parameters, the ODEs in Eq. 32 can be numerically solved. The results in this section are quite different from those in **section 3.1**. The dynamically growing compacted layer is expected to behave differently from the constant compacted layer. As time goes by, the growing compacted layer is an increasingly effective barrier to water permeation through it, which in return decelerates the system dynamics. Here, we tested four typical physical parameters ($\phi_{0,p}$, $V_{pulling}$, $R_1$, and $G$) in the model with a dynamic compacted layer to explore their possible effects on the system dynamics. The system dynamics are also characterized by the three variables $\overline{V}_i$, $\Delta \overline{T}_i$, and $\Delta \overline{Z}_i$. Here, the standard setting of physical parameters would be as follows: $V_{pulling}$ = 2 μm/s, $R_1$=1000 nm, $\phi_{0,p}$ = 0.1, and $G$ = 3 K/mm, with other parameters fixed as provided in **Table II**. We vary one of $\phi_{0,p}$, $V_{pulling}$, $R_1$, and $G$, holding the other parameters the same as those in the standard setting.

**TABLE II** Physical parameters utilized for numerical solutions in the model with a dynamic compacted layer. The parameters in a standard setting are labelled in red. The time step is taken to be 0.1 s.

| Parameters | Value | Units (SI) |
|---|---|---|
| $\phi_{0,p}$ | 0.02, 0.03, **0.1**, 0.2 | 1 |
| $V_{pulling}$ | 0.5, 1, **2**, 4 | $10^{-6}$ m/s |
| $R_1$ | 500, **1000**, 2000, 3000 | $10^{-9}$ m |
| $\phi_{I,p}$ | 0.7 | 1 |
| $\rho_1$ | 1.0 | $10^3$ kg/m$^3$ |
| $\rho_p$ | 1.1 | $10^3$ kg/m$^3$ |
| $\mu$ | 1.7921 | $10^{-3}$ Pa·s |
| $p_0$ | 1.01 | $10^5$ Pa |
| $\gamma_{S/L}$ | 35 | $10^{-3}$ J/m$^2$ |
| $L_m$ | 3.06 | $10^8$ J/m$^3$ |
| $G$ | **3**, 5, 10, 20 | $10^3$ K/m |
| $l$ | 1 | $10^{-3}$ m |

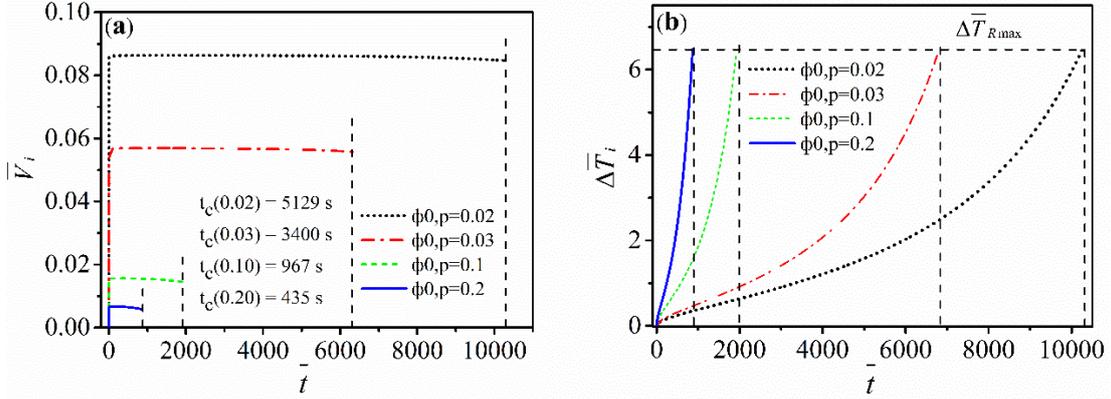

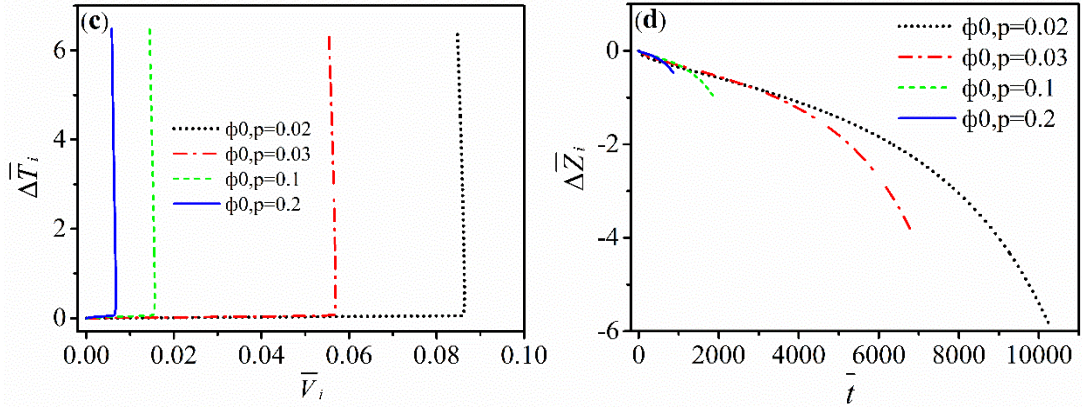

**FIG. 7** Effect of different levels of $\phi_{0,p}$ on the system dynamics with other fixed parameters from **Table II** for our model with a dynamic compacted layer. **(a)** $\bar{V}_i$-$\bar{t}$ curve with different critical times $t_c$ = 5129, 3400, 967, and 435 s for $\phi_{0,p}$ = 0.02, 0.03, 0.1, and 0.2, respectively, the vertical dashed lines indicating the physical bound by the $t_c$; **(b)** $\Delta\bar{T}_i$-$\bar{t}$ curve with labelled $\Delta\bar{T}_{R\max}$ and $t_c$ in dashed lines; **(c)** Trajectory in the ($\bar{V}_i$, $\Delta\bar{T}_i$) plane; **(d)** $\Delta\bar{Z}_i$-$\bar{t}$ curve.

**Figure 7** shows the effects of different $\phi_{0,p}$ values of 0.02, 0.03, 0.1, and 0.2 on the system dynamics with other fixed parameters in **Table II**. In the case of **Fig. 7**, $\phi_{0,p}$ controls the system dynamics in a complex manner via its relationship to $\lambda$, which, in turn, affects $F_R$, and $F_{f_D}$, as well as the nonlinear term. In **Fig. 7 (a)**, no rigorous steady state appears, which is replaced by an acceleration followed by a deceleration ending up with a specific $t_c$. The trajectories of different $\phi_{0,p}$ in the ($\bar{V}_i$, $\Delta\bar{T}_i$) plane are well separated from each other, as shown in **Fig. 7 (c)**. Larger $\phi_{0,p}$ results in faster variation of $\Delta\bar{T}_i$ and $\Delta\bar{Z}_i$, which in return shortens the $t_c$, as shown in **Fig. 7 (b)** and **Fig. 7 (d)**. For a dynamically growing compacted layer here, there is no rigorous steady state over time as that in the constant compacted layer model, as shown in **Fig. 7 (a)**. Instead, an acceleration of interface kinetics first occurs, followed by deceleration. This is not surprising because at the initial stage, permeation occurs through a very thin compacted layer that is only several times the particle radius, corresponding to a low drag force compared to the increasing pushing force. This results in an acceleration of interface kinetics similar to that in the previous constant compacted layer model. As time passes, the length of the compacted layer increases to some extent, which serves as positive feedback on the increment of the drag force. Thus, the increment of drag force of the compacted layer accelerates and increases faster than pushing force over time, since both the velocity and the length of the compacted layer increase over time. When the drag force eventually exceeds the pushing force, deceleration of interface kinetics occurs, quickly making the interface reach its critical undercooling in a certain period of time and leading to the formation of ice spears. Similarly, the whole freezing process enters the regime of ice spear growth when $t_c$ is reached, which is beyond the scope of our model and serves as a *physical bound* for $\bar{V}_i$, $\Delta \bar{T}_i$ and $\Delta \bar{Z}_i$. The calculated data points for all quantities beyond the range of critical time are thus *invalid* and not included in our results.

It is worth noting that a kinetic plateau of finite length and magnitude is observed in numerical solutions under some physical conditions, which is somewhat similar to the observed steady state in the constant compacted layer model. That is, under some choice of parameters, the interface kinetics can keep up with the pulling velocity in a finite period of time but are still slightly lower than the pulling velocity. Within this plateau, the deceleration process is relaxed, whose variation is not dramatic until after a finite duration of time. This "kinetic plateau" is usually observed for compacted layers with higher permeability (i.e., lower $\phi_{0,p}$, as shown in **Fig. 7 (a)**, and larger $R_1$, as shown in **Fig. 9 (a)**) and under lower $V_{pulling}$, as shown in **Fig. 8 (a)**. For the compacted layers with higher permeability and under lower pulling velocity, the drag force is not very significant at the initial stage of freezing, leading to the ability of interface kinetics to quickly keep up with the pulling velocity and result in a "kinetic plateau". Within this plateau, deceleration of the interface kinetics is very slow because the permeation thickness is too thin to result in a significant drag force at the initial stage, and the increasing pushing force can still keep up with the increasing drag force. As the thickness of the compacted layer further increases to some extent over time, the drag force increases faster and faster, which eventually surpasses the pushing force and results in significant deceleration of interface kinetics, eliminating the sustainment of the "kinetic plateau". On the other hand, for the compacted layers with lower permeability (i.e., lower $\phi_{0,p}$, as shown in **Fig. 7 (a)**, and smaller $R_1$, as shown in **Fig. 9 (a)**), and under higher $V_{pulling}$, as shown in **Fig. 8 (a)**, the "kinetic plateau" is shortened and even disappears.

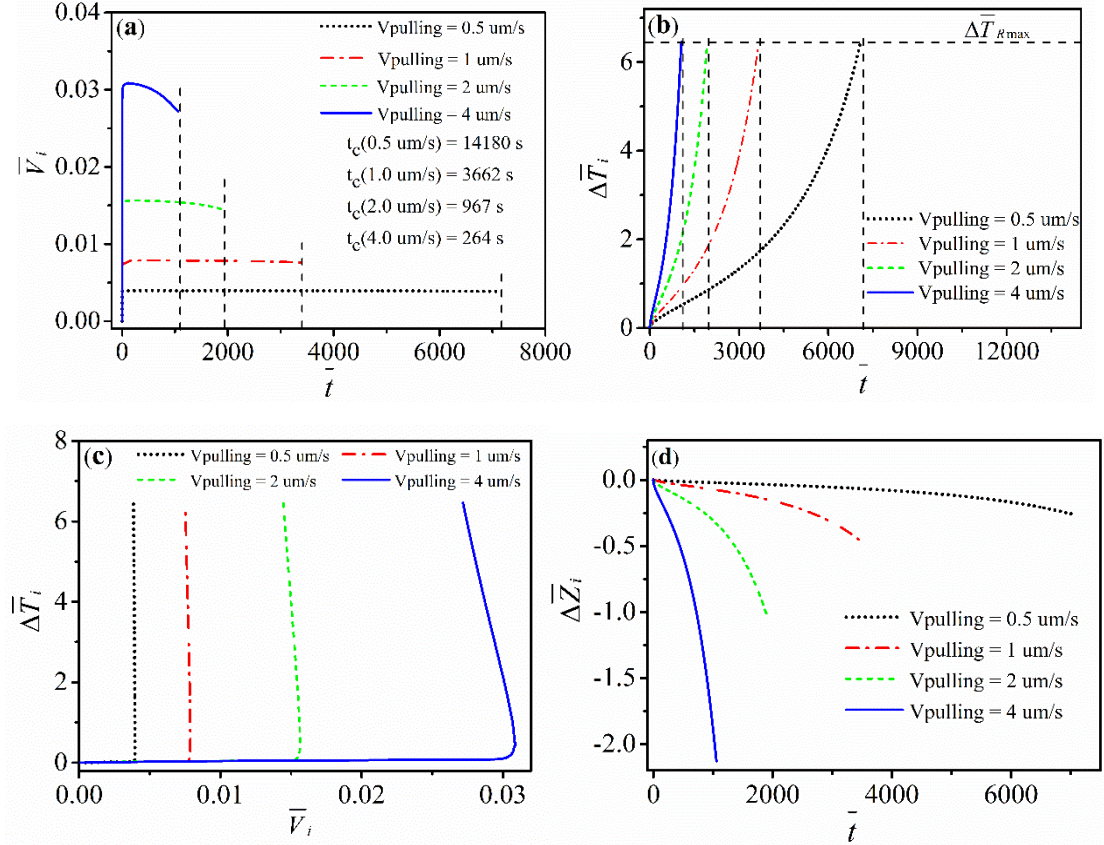

**FIG. 8** Effects of different levels of $V_{pulling}$ on the system dynamics with other fixed parameters from **Table II** for our model with a dynamic compacted layer. **(a)** $\bar{V}_i$-$\bar{t}$ curve with different critical times $t_c$ = 14180, 3662, 967, and 264 s for $V_{pulling}$ = 0.5, 1, 2, and 4 μm/s, respectively, the vertical dashed lines indicate the physical bound by the $t_c$; **(b)** $\Delta\bar{T}_i$-$\bar{t}$ curve with labelled $\Delta\bar{T}_{R\max}$ and $t_c$ in dashed lines; **(c)** Trajectory in the ($\bar{V}_i$, $\Delta\bar{T}_i$) plane; **(d)** $\Delta\bar{Z}_i$-$\bar{t}$ curve.

**Figure 8** shows the effect of different $V_{pulling}$ values of 0.5, 1, 2, and 4 μm/s on the system dynamics with other fixed parameters in **Table II**. In the case of **Fig. 8**, $V_{pulling}$ seems to control the system dynamics in a simple manner via its relation with $\Delta \bar{T}_i$, which affects $F_R$. In **Fig. 8 (a)**, an acceleration followed by a deceleration ending up with a specific $t_c$ is also observed. The trajectories of different $V_{pulling}$ in the ($\bar{V}_i$, $\Delta \bar{T}_i$) plane are well separated from each other, as shown in **Fig. 8 (c)**. In addition, similar to the effects of increased $\phi_{0,p}$, larger $V_{pulling}$ also results in faster variation of $\Delta \bar{T}_i$ and $\Delta \bar{Z}_i$, which in return shortens the $t_c$, as shown in **Fig. 8 (b)** and **Fig. 8 (d)**.

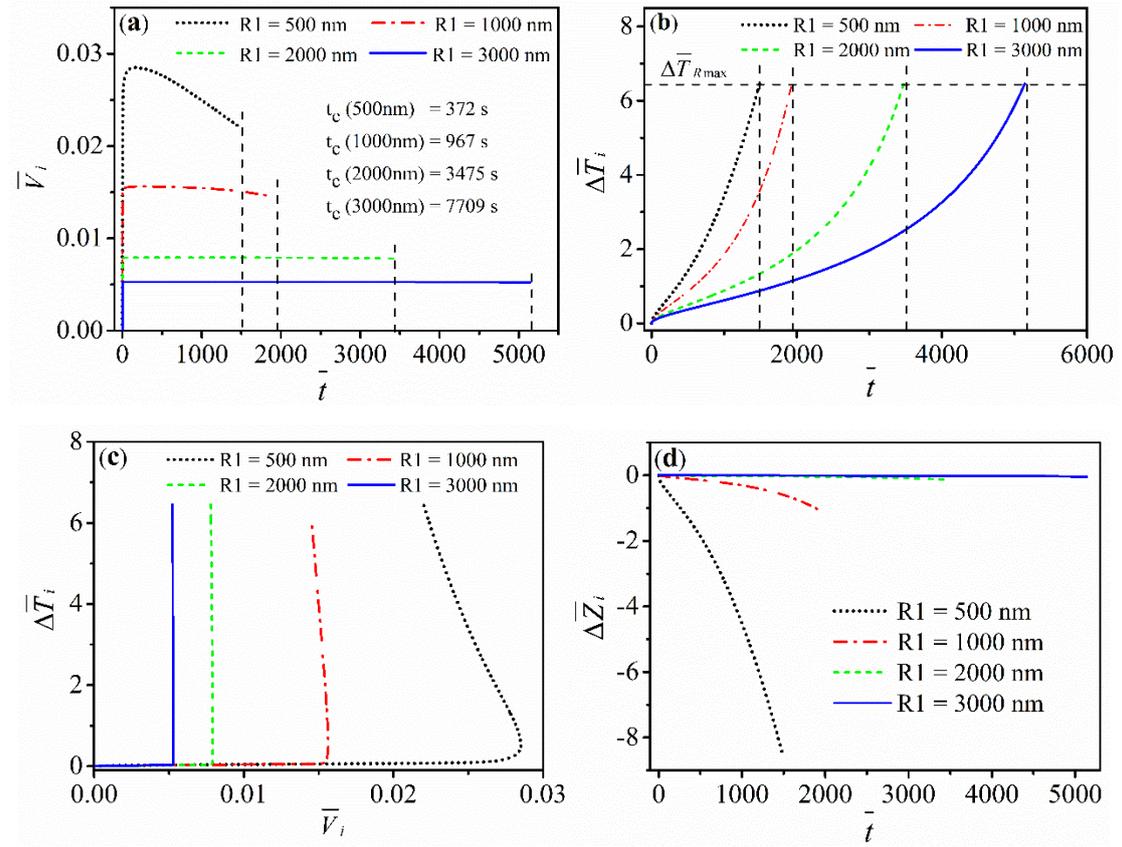

**FIG. 9** Effects of different levels of $R_1$ on system dynamics with other fixed parameters from **Table II** for our model with a dynamic compacted layer. **(a)** $\bar{V}_i$-$\bar{t}$ curve with different critical times $t_c$ = 372, 967, 3475, and 7709 s for $R_1$ = 500, 1000, 2000, and 3000 nm, respectively, the vertical dashed lines indicate the physical bound by the $t_c$; **(b)** $\Delta \bar{T}_i$-$\bar{t}$ curve with labelled $\Delta \bar{T}_{R\max}$ and $t_c$ in dashed lines; **(c)** Trajectory in the ($\bar{V}_i$, $\Delta \bar{T}_i$) plane; **(d)** $\Delta \bar{Z}_i$-$\bar{t}$ curve.

**Figure 9** shows the effect of different $R_1$ values of 500, 1000, 2000, and 3000 nm on the system dynamics with other fixed parameters in **Table II**. In the case of **Fig. 9**, $R_1$ controls the system dynamics in a manner more complex than $V_{pulling}$, which affects both $F_{f_D}$ and $F_R$. In **Fig. 9 (a)**, an acceleration followed by a deceleration ending up with a specific $t_c$ is also observed. The trajectories of different $R_1$ in the ($\bar{V}_i$, $\Delta\bar{T}_i$) plane are well separated from each other, as shown in **Fig. 9 (c)**. In addition, unlike the effects of increased $\phi_{0,p}$ and/or $V_{pulling}$, larger $R_1$ results in slower variation of $\Delta\bar{T}_i$ and $\Delta\bar{Z}_i$, together with lower $\Delta\bar{T}_{R\max}$, which, nevertheless, enlarges the $t_c$, as shown in **Fig. 9 (b)** and **Fig. 9 (d)**.

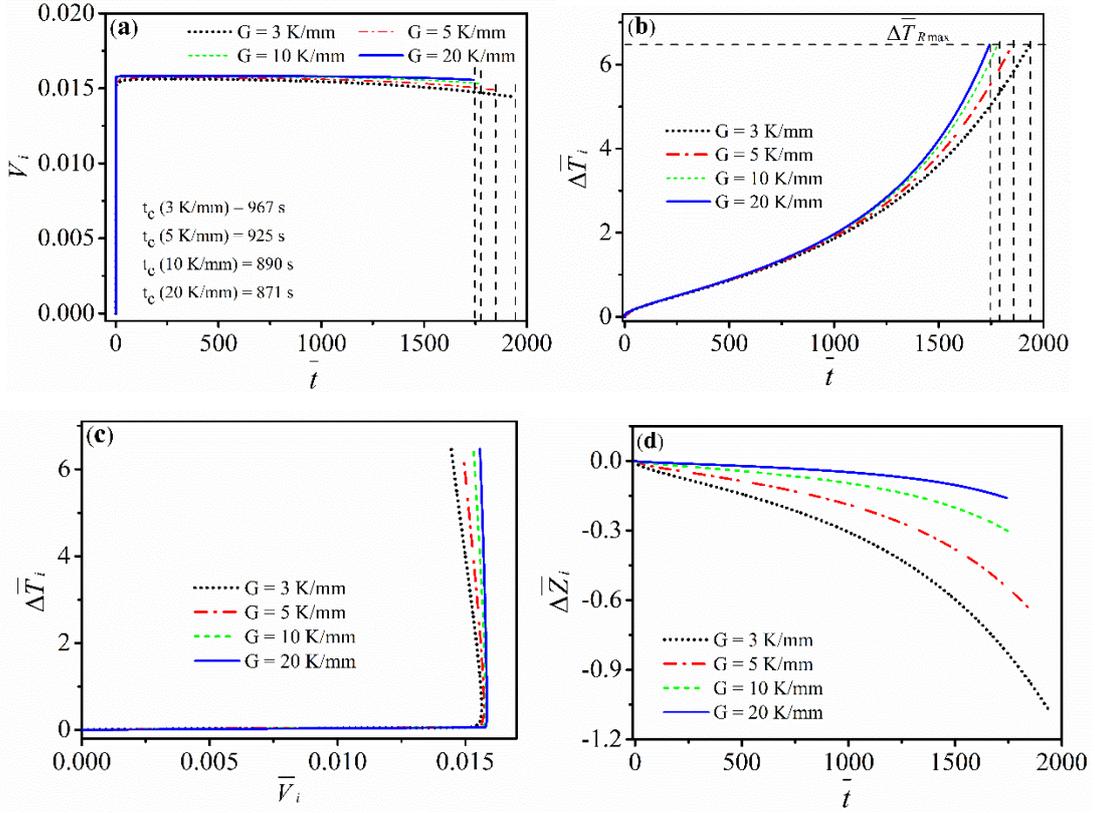

**FIG. 10** Effects of different levels of $G$ on the system dynamics with other fixed parameters from **Table II** for our model with a dynamic compacted layer. **(a)** $\bar{V}_i$-$\bar{t}$ curve with different critical times $t_c$ = 967, 925, 890, and 871 s for $G$ = 3, 5, 10, and 20 K/mm, respectively, the vertical dashed lines indicating the physical bound by the $t_c$; **(b)** $\Delta\bar{T}_i$-$\bar{t}$ curve with labelled $\Delta\bar{T}_{R\max}$ and $t_c$ in dashed lines; **(c)** Trajectory in the ($\bar{V}_i$, $\Delta\bar{T}_i$) plane; **(d)** $\Delta\bar{Z}_i$-$\bar{t}$ curve.

**Figure 10** shows the effects of different $G$ values of 3, 5, 10, and 20 K/mm on the system dynamics in relation to other fixed parameters in **Table II**. In the case of **Fig. 10**, $G$ controls the system dynamics in a simple manner via its relationship with $\Delta \bar{T_i}$, which affects $F_R$. In **Fig. 10 (a)**, an acceleration followed by a deceleration ending up with a specific $t_c$ is also observed, which shows that increased $G$ increases the length of the "kinetic plateau". The trajectories of different $R_1$ in the ($\bar{V_i}$, $\Delta \bar{T_i}$) plane are well separated from each other, as shown in **Fig. 10 (c)**. In addition, similar to the effects of increased $\phi_{0,p}$ and/or $V_{pulling}$, larger $G$ results in faster variation of $\Delta \bar{T_i}$ and $\Delta \bar{Z_i}$, which in return shortens the $t_c$, as shown in **Fig. 10 (b)** and **Fig. 10 (d)**.

## IV. CONCLUSION

This paper proposes a theoretical framework based on the momentum theorem of a constant/dynamic compacted layer as a consequence of the two interacting forces $F_R$ and $F_{f_D}$ to address the nonlinear dynamic behavior of the unidirectional freezing process of particle suspensions. Many typical physical parameters are incorporated into the models in this paper, which can be finely tuned to alter the dynamics of the unidirectional freezing of particle suspensions.

For a constant compacted layer with higher permeability (i.e., larger $R_1$ and smaller $L_0$) and under lower $V_{pulling}$, the interface kinetics are more likely to experience a physically possible steady state. In contrast, for a constant compacted layer with lower permeability and under higher $V_{pulling}$, it will be more difficult for the interface kinetics to keep up with the $V_{pulling}$ before the $t_c$ is reached to reach a physically possible steady state. For a dynamically growing compacted layer, the interface kinetics first undergoes a steep acceleration, which is followed by deceleration. It is important to note that the predicted growth kinetics in our model are similar to previous predictions [46], in which the growth kinetics of ice increases steeply as a function of interface undercooling and decreases afterwards. A "kinetic plateau" of finite length and magnitude is found by numerical calculation from different choices of physical parameters, in which the deceleration of interface kinetics is not dramatic until after a finite duration of time. It is shown that the interface kinetics are more likely to experience a "kinetic plateau" with higher permeability (i.e., larger $R_1$, lower $\phi_{0,p}$) and under lower $V_{pulling}$. In contrast, for a dynamically growing compacted layer with lower permeability under larger $V_{pulling}$, the "kinetic plateau" is shortened and even disappears.

The theoretical framework proposed in this paper allows us to reconsider the mechanisms of pattern formation in the unidirectional freezing of particle suspensions in a simple but novel way, with potential implications for both understanding and controlling not only ice formations in porous media but also crystallization processes in other complex systems. Further explorations are needed to describe the unidirectional freezing of more complex systems, such as aqueous solutions of macromolecules [26, 47, 48], in which the interaction between water permeation through a porous network and build-up of a solute diffusion-controlled boundary layer are to be considered. In addition, the microstructure of the porous media also needs to be considered in the future, as an inhomogeneous microstructure [49] can greatly affect the permeation process.

**ACKNOWLEDGEMENTS**

The work was supported by the Research Fund of the State Key Laboratory of Solidification Processing (NPU), China (Grant No. 2020-TS-06, 2021-TS-02).

**DATA AVAILABILITY**

The data that support the findings of this study are available from the corresponding author upon reasonable request.

**Appendix A: Comparison of the models with a constant compacted layer with and without the term $d\bar{V}_i / d\bar{t}$**

In this section, we compare the validity of the simplification, which assumes that the term $d\bar{V}_i / d\bar{t}$ in Eq. 29 is very small and negligible. The model based on this assumption is termed the "inertia-free" model in the following. When we set $d\bar{V}_i / d\bar{t}$ to zero as a force balance, the relation between $\bar{V}_i$ and $\Delta \bar{T}_i$ can be obtained

$$\bar{V}_i = \left(1 - \frac{1}{1+\Delta \bar{T}_i}\right)^2 \qquad \text{(Eq. A1)}$$

Substituting Eq. A1 into Eq. 29 to replace $\bar{V}_i$, we obtain a single component ODE about $\Delta \bar{T}_i$ as

$$\frac{d\Delta \bar{T}_i}{d\bar{t}} = C_2 \left[\frac{V_{pulling}}{[V]} - \left(1 - \frac{1}{1+\Delta \bar{T}_i}\right)^2\right] \qquad \text{(Eq. A2)}$$

with the initial condition

$$\Delta \bar{T}_i(\bar{t}=0) = 0 \qquad \text{(Eq. A3)}$$

By separations of variables, we have

$$\frac{d\Delta \bar{T}_i}{\frac{V_{pulling}}{[V]} - \left(1 - \frac{1}{1+\Delta \bar{T}_i}\right)^2} = C_2 d\bar{t} \qquad \text{(Eq. A4)}$$

Integration on both sides of Eq. A4 in combination with Eq. A3 gives the exact solution of $\Delta \bar{T}_i$ for the "inertia-free" model, which is a very complex implicit function $f(\Delta \bar{T}_i, \bar{t}) = 0$ as

$$C_2 \bar{t} + \frac{\ln \bar{V}_{pl}}{2(\bar{V}_{pl}+1)\sqrt{\bar{V}_{pl}}} + \frac{\Delta \bar{T}_i}{1-\bar{V}_{pl}} + \frac{2\bar{V}_{pl} \ln\left[\frac{(\Delta \bar{T}_i+1)\sqrt{\bar{V}_{pl}} + \Delta \bar{T}_i}{(\Delta \bar{T}_i+1)\sqrt{\bar{V}_{pl}} - \Delta \bar{T}_i}\right] - (\bar{V}_{pl}+1)\sqrt{\bar{V}_{pl}} \ln[\bar{V}_{pl}(\Delta \bar{T}_i+1)^2 - \Delta \bar{T}_i^2]}{2\bar{V}_{pl}(\bar{V}_{pl}+1)^2} = 0 \qquad \text{(Eq. A5)}$$

where $\bar{V}_{pl} = V_{pulling} / [V]$ is defined as the nondimensional form of $V_{pulling}$.

For simplicity, we use the numerical solution of Eq. A2 for comparison. By varying the particle radius (1000 and 300 nm) and pulling velocity (2 and 6 um/s), we plot the results of the two models in **Fig. A1**. The results of the "inertia-free" model are very close to those of the original model, which validates the simplification of zero acceleration for the case of a constant compacted layer. However, we use numerical solutions of Eq. 29 for analysis and discussion in the main body of the paper.

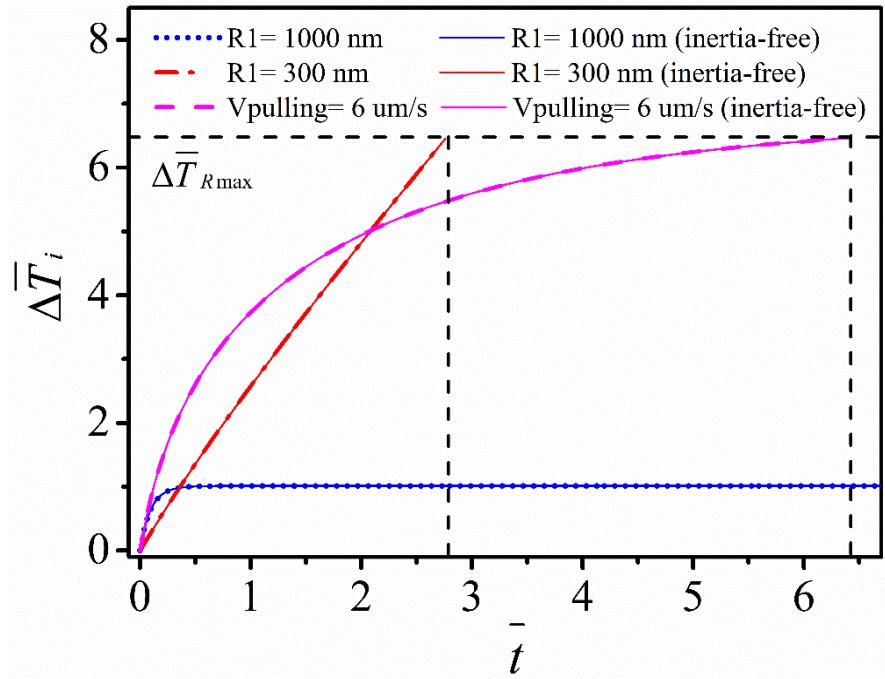

**Fig. A1** Comparison of the prediction of $\Delta \bar{T}_i$ as a function of $\bar{t}$ between the constant layer models with and without inertia ("inertia-free" model) of the particles.

**Appendix B: Comparison of the models with experimental observations**

To our knowledge so far, a few observations have reported ice growth kinetics in particle suspensions [25, 50, 51]. However, directly comparable literature is very rare because the investigated system in this paper is a pure ice-water system, which must exclude the solutal effect. The presence of dissolved additives in particle suspensions can make the growth kinetics of ice much more complex, as solute diffusion in porous media is tricky and not involved in the present paper. We found that the literature observing ice lenses growth by Schollick et al. [51] is probably most comparable to our model. It is indicated that the packing density is not necessarily a fixed value [51]. The data points of reference [51] in **Fig. B1** are estimated by the averaged growth velocity and the time of a growing ice lens before the emergence of a new ice lens. Thus, by choosing an appropriate packing density, we found that our model can well match the reported growth rate of an individual ice lens in particle suspensions free of impurities [51]. The discrepancy in critical time between the literature [51] and our model may result from the inhomogeneity of the packing state in the direction normal to the pulling velocity, as well as the polydispersity of the particles used, resulting in earlier formation of a new ice lens.

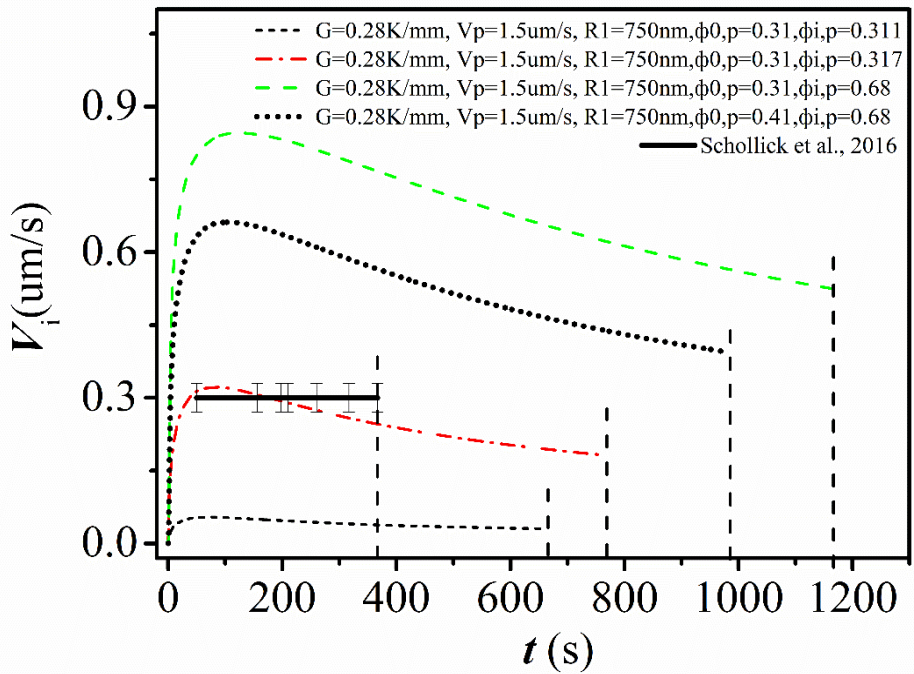

**Fig. B1** Comparison of the prediction of ice growth kinetics $V_i$ as a function of time $t$ between the predicted results by our dynamically growing layer model with varying packing density and the experimental observations of ice lenses growth by Schollick et al. [51].

Captions for all Figures and Tables:

**FIG. 1** Schematic of two cases of unidirectional freezing. The freezing sample is assumed to be static in the frame of reference $Z$, which is fixed to the ground. The isotherm of $T_m$ with its position $Z_{T_m}(t)$ in a thermal gradient of fixed magnitude $G$ moves horizontally from left to right at speed $V_{pulling}$, leading to a unidirectional freezing of water in the sample. As the ice grows forwards, the constant compacted layer with a constant packing density $\phi_{I,p}$ of particles is pushed forwards ahead of the ice due to force differences ($F_R$ and $F_{f_D}$) exerted on it to result in a corresponding interface movement $Z_i(t)$. The ice growth is supported by water inflow at a velocity $u$ from right to left through the compacted layer. **(a)** A compacted layer of constant length $L_0$ is placed above an ice/water interface in a glass capillary that is glued onto a large glass sheet, forming two interfaces (the solid/particle interface $\Gamma_{S/P}$ and the particle/liquid interface $\Gamma_{P/L}$) with no particle far from the compacted layer. **(b)** A particle suspension is placed above an ice/water interface in a glass capillary glued onto a large glass sheet. As the ice grows forwards, the particles accumulate ahead of the ice, which results in a dynamically growing compacted layer of particles with a time-dependent length $L(t)$ and a constant packing density of particles $\phi_{I,p}$, forming two interfaces ($\Gamma_{S/P}$ and $\Gamma_{P/L}$).

**FIG. 2 (a)** Schematic for the formation of curvature radii for indentation ($R_1$) and pore ice ($R_2$) protruding into the porous interstices of the compacted particles on $\Gamma_{S/P}$ due to its contact with particles with a time-dependent contact angle of $\theta$, which is assumed to satisfy Eq. 6. **(b)** Schematic for the formation of a pushing force exerted on the compacted layer due to surface tension at the ice/water/particle interface with a time-dependent tip radius $R_2(t)$ of pore ice between porous interstices near $\Gamma_{S/P}$, in which $t_1 < t_2 < t_3$, and $R_2(t_1) > R_2(t_2) > R_2(t_3) \geq R_1$.

**FIG. 3** Effects of different levels of $V_{pulling}$ on system dynamics with other fixed parameters from **Table I** for our model with a constant compacted layer. **(a)** $\bar{V}_i$-$\bar{t}$

curve with different $t_c$ = 108, and 20 s for $V_{pulling}$ = 6, and 8 μm/s, respectively, the $t_c$ for $V_{pulling}$ = 1, and 2 μm/s being infinite due to the appearance of their steady state before reaching their $\Delta \bar{T}_{R\max}$, the vertical dashed lines indicating the physical bound by the $t_c$; **(b)** $\Delta \bar{T}_i$ - $\bar{t}$ curve with labelled $\Delta \bar{T}_{R\max}$ and $t_c$ in dashed lines; **(c)** Trajectory in the ($\bar{V}_i$, $\Delta \bar{T}_i$) plane; **(d)** $\Delta \bar{Z}_i$ - $\bar{t}$ curve.

**FIG. 4** Effects of different levels of $R_1$ on system dynamics with other fixed parameters from **Table I** for our model with a constant compacted layer. **(a)** $\bar{V}_i$ - $\bar{t}$ curve with different critical times $t_c$ = 138, and 159 s for $R_1$ = 300, and 500 nm, respectively, the $t_c$ for $R_1$ = 1000, and 3000 nm being infinite due to the appearance of their steady state before reaching their $\Delta \bar{T}_{R\max}$, the vertical dashed lines indicating the physical bound by the $t_c$; **(b)** $\Delta \bar{T}_i$ - $\bar{t}$ curve with labelled $\Delta \bar{T}_{R\max}$ and $t_c$ in dashed lines; **(c)** Trajectory in the ($\bar{V}_i$, $\Delta \bar{T}_i$) plane; **(d)** $\Delta \bar{Z}_i$ - $\bar{t}$ curve.

**FIG. 5** Effect of different levels of $L_0$ on the system dynamics with other fixed parameters from **Table I** for our model with a constant compacted layer. **(a)** $V_i(t)$ - $t$ curve with roughly the same critical time $t_c$ = 80, and 320 s for $L_0$ = 300, and 400 μm, respectively, the $t_c$ for $L_0$ = 10, and 100 μm being infinite due to the appearance of their steady state before reaching their $\Delta T_{R\max}$, the vertical dashed lines indicating the physical bound by the $t_c$; **(b)** $\Delta T_i(t)$ - $t$ curve with labelled $\Delta T_{R\max}$ and $t_c$ in dashed lines; **(c)** Trajectory in the ($V_i(t)$, $\Delta T_i(t)$) plane; **(d)** $\Delta Z_i(t)$ - $t$ curve.

**FIG. 6** Effects of different levels of $G$ on the system dynamics with other fixed

parameters from **Table I** for our model with a constant compacted layer. **(a)** $\overline{V}_i$-$\overline{t}$ curve with infinite critical times $t_c$ for $G$ = 1, 3, 5, and 7 K/mm due to the appearance of their steady state before reaching their $\Delta \overline{T}_{R\max}$; **(b)** $\Delta \overline{T}_i$-$\overline{t}$ curve with labelled $\Delta \overline{T}_{R\max}$ in dashed line; **(c)** Trajectory in the ($\overline{V}_i$, $\Delta \overline{T}_i$) plane; **(d)** $\Delta \overline{Z}_i$-$\overline{t}$ curve.

**FIG. 7** Effect of different levels of $\phi_{0,p}$ on the system dynamics with other fixed parameters from **Table II** for our model with a dynamic compacted layer. **(a)** $\overline{V}_i$-$\overline{t}$ curve with different critical times $t_c$ = 5129, 3400, 967, and 435 s for $\phi_{0,p}$ = 0.02, 0.03, 0.1, and 0.2, respectively, the vertical dashed lines indicating the physical bound by the $t_c$; **(b)** $\Delta \overline{T}_i$-$\overline{t}$ curve with labelled $\Delta \overline{T}_{R\max}$ and $t_c$ in dashed lines; **(c)** Trajectory in the ($\overline{V}_i$, $\Delta \overline{T}_i$) plane; **(d)** $\Delta \overline{Z}_i$-$\overline{t}$ curve.

**FIG. 8** Effects of different levels of $V_{pulling}$ on the system dynamics with other fixed parameters from **Table II** for our model with a dynamic compacted layer. **(a)** $\overline{V}_i$-$\overline{t}$ curve with different critical times $t_c$ = 14180, 3662, 967, and 264 s for $V_{pulling}$ = 0.5, 1, 2, and 4 μm/s, respectively, the vertical dashed lines indicate the physical bound by the $t_c$; **(b)** $\Delta \overline{T}_i$-$\overline{t}$ curve with labelled $\Delta \overline{T}_{R\max}$ and $t_c$ in dashed lines; **(c)** Trajectory in the ($\overline{V}_i$, $\Delta \overline{T}_i$) plane; **(d)** $\Delta \overline{Z}_i$-$\overline{t}$ curve.

**FIG. 9** Effects of different levels of $R_1$ on system dynamics with other fixed parameters from **Table II** for our model with a dynamic compacted layer. **(a)** $\overline{V}_i$-$\overline{t}$ curve with different critical times $t_c$ = 372, 967, 3475, and 7709 s for $R_1$ = 500, 1000, 2000, and 3000 nm, respectively, the vertical dashed lines indicate the physical bound

by the $t_c$ ; **(b)** $\Delta \bar{T_i}$ - $\bar{t}$ curve with labelled $\Delta \bar{T}_{R\max}$ and $t_c$ in dashed lines; **(c)** Trajectory in the ($\bar{V_i}$, $\Delta \bar{T_i}$) plane; **(d)** $\Delta \bar{Z_i}$ - $\bar{t}$ curve.

**FIG. 10** Effects of different levels of $G$ on the system dynamics with other fixed parameters from **Table II** for our model with a dynamic compacted layer. **(a)** $\bar{V_i}$ - $\bar{t}$ curve with different critical times $t_c$ = 967, 925, 890, and 871 s for $G$ = 3, 5, 10, and 20 K/mm, respectively, the vertical dashed lines indicating the physical bound by the $t_c$; **(b)** $\Delta \bar{T_i}$ - $\bar{t}$ curve with labelled $\Delta \bar{T}_{R\max}$ and $t_c$ in dashed lines; **(c)** Trajectory in the ($\bar{V_i}$, $\Delta \bar{T_i}$) plane; **(d)** $\Delta \bar{Z_i}$ - $\bar{t}$ curve.

**Fig. A1** Comparison of the prediction of $\Delta \bar{T_i}$ as a function of $\bar{t}$ between the constant layer models with and without inertia ("inertia-free" model) of the particles.

**Fig. B1** Comparison of the prediction of ice growth kinetics $V_i$ as a function of time $t$ between the predicted results by our dynamically growing layer model with varying packing density and the experimental observations of ice lenses growth by Schollick et al. [51].

**TABLE I** Physical parameters utilized for numerical solutions in the model with a constant compacted layer. The parameters in a standard setting are labelled in red. The time step is taken to be 0.1 s.

**TABLE II** Physical parameters utilized for numerical solutions in the model with a dynamic compacted layer. The parameters in a standard setting are labelled in red.

The time step is taken to be 0.1 s.